\documentclass[aps,prd,superscriptaddress,nofootinbib,tighten,preprint]{revtex4}

\newcommand{\be}{\begin{eqnarray}}
\newcommand{\ee}{\end{eqnarray}}

\def\numu{{\nu_{\mu}}}

\newcommand{\ms}{\Delta m^2_{21}}
\newcommand{\ma}{\Delta m^2_{31}}

\def\nn{\nonumber}

\newcommand{\epsem}{\epsilon_{e\mu}}
\newcommand{\epset}{\epsilon_{e\tau}}

\newcommand{\epsmt}{\epsilon_{\mu\tau}}
\newcommand{\epstt}{\epsilon_{\tau\tau}}

\def\gs{\mathrel{
   \rlap{\raise 0.511ex \hbox{$>$}}{\lower 0.511ex \hbox{$\sim$}}}}
\def\ls{\mathrel{
   \rlap{\raise 0.511ex \hbox{$<$}}{\lower 0.511ex \hbox{$\sim$}}}}
\newcommand{\bea}{\begin{equation} \begin{array}{c}}
\newcommand{\bead}{\begin{equation} \begin{array}{cccc}}
\newcommand{\eea}{ \end{array} \end{equation}}
\usepackage{amsfonts}
\usepackage{amssymb}
\usepackage{amsmath}
\usepackage{graphicx,subfigure}
\usepackage{bbm}
\def\slc#1{\setbox0=\hbox{$#1$}           
    \dimen0=\wd0                                 
    \setbox1=\hbox{/} \dimen1=\wd1               
    \ifdim\dimen0>\dimen1                        
       \rlap{\hbox to \dimen0{\hfil/\hfil}}      
       #1                                        
    \else                                        
       \rlap{\hbox to \dimen1{\hfil$#1$\hfil}}   
       /                                         
    \fi}

\newcommand{\eg}{\emph{e.g.}}
\newcommand{\ie}{\emph{i.e.}}

\begin{document}

\title{Neutrino Physics with Non-Standard Interactions at INO}

\author{Sandhya Choubey}
\email{sandhya@hri.res.in}
\affiliation{Harish-Chandra Research Institute, Chhatnag Road, Jhunsi, Allahabad 211 019, India}
\affiliation{Department of Theoretical Physics, School of Engineering Sciences, KTH Royal Institute of Technology, AlbaNova University Center, 106 91 Stockholm, Sweden}

\author{Anushree Ghosh}
\email{anushree@cbpf.br}
\affiliation{Centro Brasileiro de Pesquisas F{\'i}sicas, Rua Dr.~Xavier Sigaud 150, Urca, Rio de Janeiro, RJ, 22290-180, Brazil}

\author{Tommy Ohlsson}
\email{tohlsson@kth.se}
\affiliation{Department of Theoretical Physics, School of Engineering Sciences, KTH Royal Institute of Technology, AlbaNova University Center, 106 91 Stockholm, Sweden}

\author{Deepak Tiwari}
\email{deepaktiwari@hri.res.in}
\affiliation{Harish-Chandra Research Institute, Chhatnag Road, Jhunsi, Allahabad 211 019, India}

\begin{abstract}
Non-standard neutrino interactions (NSI) involved in neutrino propagation 
inside Earth matter could potentially alter atmospheric neutrino fluxes. In this 
work, we look at the impact of these NSI on the signal at the ICAL detector 
to be built at the India-based Neutrino Observatory (INO). 
We show how 
the sensitivity to the neutrino mass hierarchy of ICAL changes in the presence of 
NSI. The mass hierarchy sensitivity is shown to be rather sensitive to 
the NSI parameters $\epsilon_{e\mu}$ and $\epsilon_{e\tau}$, while 
the dependence on $\epsilon_{\mu\tau}$ and $\epsilon_{\tau\tau}$ is seen to 
be very mild, once the $\chi^2$ is marginalised over oscillation and NSI parameters. 
If the NSI are large enough, the event spectrum at ICAL is expected to be altered 
and this can be used to discover new physics. 
We calculate the lower limit on NSI parameters above which 
ICAL could discover NSI at a given C.L.~from 10 years of data. If NSI were too small, 
the null signal at ICAL can constrain the NSI parameters. We give  
upper limits on the NSI parameters at any given C.L.~that one is expected 
to put from 10 years of running of ICAL.  Finally, we give C.L.~contours 
in the NSI parameter space that is expected to be still allowed from 10 years of 
running of the experiment.

\end{abstract}

\maketitle

\section{\label{intro} Introduction}

The 50 kton magnetised iron detector (ICAL) to be built at the India-based neutrino 
observatory (INO) will be mainly observing muon neutrinos coming from the Earth's 
atmosphere \cite{Ahmed:2015jtv}. 
Amongst the most important goals of this experiment is the determination 
of the neutrino mass hierarchy (MH) 
through the observation of Earth matter effects in the 
expected data sample. For the mass-squared difference $\ma > 0$ 
(normal hierarchy)\footnote{We define $\Delta m_{ij}^2 = m_i^2-m_j^2$.}, 
one expects matter enhanced oscillations in the neutrino channel in the 
energy range around (5-10) GeV, while the antineutrino channel does 
not experience any such 
matter induced enhancement. On the other hand for 
$\ma < 0$ (inverted hierarchy), 
matter enhanced oscillations are expected in the antineutrino channel, 
while the neutrino channel does not obtain any such enhancement. 
ICAL being magnetised will be able to tell its 
neutrino signal from its antineutrino signal, 
giving the detector an added handle on measuring these 
Earth matter effects, and hence, the 
neutrino mass hierarchy. The sensitivity reach of this experiment for measuring 
standard neutrino oscillation parameters have been studied extensively in 
Refs.~\cite{Ghosh:2012px,Thakore:2013xqa,Ghosh:2013mga,Devi:2014yaa,Kaur:2014rxa}. 

So far, there has been no signal of physics beyond the Standard Model 
in any of the accelerator-based 
experiments including LHC. However, we have unambiguous evidence from 
complimentary experiments 
that the Glashow--Weinberg--Salam 
model of elementary particles is at best a low-energy effective 
theory and that there exists physics beyond the Standard Model. 
The Standard Model of particle physics is unable to convincingly 
explain data from neutrino oscillation experiments. In addition, it also fails  
to provide explanation of the existence of dark matter 
and dark energy in the Universe, as well as baryogenesis. There are 
also theoretical 
issues with the Standard Model which demand its extension. Any extension of the 
model could include addition to its gauge or particle sector, or both. 
It is therefore pertinent to envisage that such an extended theory would also 
have new (effective) interactions between the particles, beyond what is included in the 
Standard Model. Such interactions are expected to change the 
predicted outcome of experiments and existing data can be used to put limits 
on the strength of these interactions. In this work, we are primarily 
interested in non-standard interactions (NSI) affecting neutrino oscillations 
as neutrinos propagate inside Earth matter. These NSI, if present, would modify the 
transport of atmospheric neutrinos  inside Earth 
matter, and hence alter the signal at the ICAL detector. 

The currently running Super-Kamiokande (SK) atmospheric neutrino experiment 
has looked for possible presence of these NSI in its event sample, and 
has found the data 
to be completely consistent with the Standard Model. Through a statistical analysis, 
the SK collaboration converts 
this into an upper bound on the relevant NSI parameters \cite{Mitsuka:2011ty}. 
Expected constraints from other (future) atmospheric neutrino 
experiments have been studied 
previously in the literature, see 
\eg~Refs.~\cite{Ohlsson:2013epa,Esmaili:2013fva,Choubey:2014iia,Mocioiu:2014gua,Chatterjee:2014gxa,Yasuda:2015lwa} (see Ref.~\cite{Fornengo:2001pm,GonzalezGarcia:2004wg,Friedland:2004ah,Friedland:2005vy,GonzalezGarcia:2011my,Escrihuela:2011cf} for earlier works). 

In this paper, we will study in detail the impact of NSI on the atmospheric neutrino 
signal in the ICAL detector at INO. We analyse the prospective data at ICAL in terms of 
the measured muon energy and muon angle through a binned $\chi^2$ analysis. 
ICAL is expected to also measure the energy deposited in the associated hadron 
shower. Inclusion of the hadron energy information improves the energy 
reconstruction of the events and hence in general improves the sensitivity of 
ICAL  \cite{Devi:2014yaa}. We have not included the hadron energy information in this work. 
This is being studied in a follow-up work by the INO collaboration \cite{nsihadron}.  
We  
use the Nuance event generator with the ICAL detector geometry for 
generating muons from atmospheric neutrinos. The ICAL energy and angle resolutions and 
reconstruction and charge identification efficiencies are obtained from 
the Geant-based detector simulation code developed for ICAL. We  
generate muon events in the range (1-100)~GeV and show the increase in the 
sensitivity to NSI parameters with the increase of the neutrino energy, and hence the muon 
energy, as was pointed out in Ref.~\cite{Choubey:2014iia}. 
We will quantify the extent of this 
modification in the expected muon signal at ICAL. Using that we will study 
the expected limits that ICAL could impose on NSI parameters if there is 
no evidence of NSI in the data. If on the other hand the NSI parameters are 
large enough, we would see a signal of new physics in the ICAL data. 
We give the lower limit on the NSI parameters which is needed in order to 
allow their discovery in ICAL at any given C.L. 
Likewise, the presence of NSI could 
change the sensitivity of ICAL to other neutrino oscillation parameters. In 
particular, we will show how the NSI parameters alter the mass hierarchy sensitivity 
in this class of experiments and present the revised sensitivity limits. 

The paper is organised as follows. In Section \ref{prob}, we discuss the 
neutrino oscillation probabilities in the presence of NSI. In particular, we 
study the impact of NSI on the difference in the probabilities between the NH and 
IH cases. In Section \ref{events}, we describe the ICAL detector, our 
simulation techniques, and the statistical analysis procedure. We present our 
main results in Section \ref{results}. All results are shown for 500 kton-year of 
data in ICAL. 
Subsection \ref{MH} is devoted to study the impact of NSI on the mass hierarchy sensitivity 
of ICAL. In Subsections \ref{bounds}, \ref{discovery}, and \ref{precision}, we discuss   
the expected constraints on NSI parameters, the expected lower limit allowing for 
discovery of NSI, and the allowed areas in NSI parameters space, respectively. 
We end in Section \ref{conclusions} with our conclusions.

\section{\label{prob}Impact of NSI on Oscillation Probabilities}

As outlined in the Introduction, an extension of the Standard Model of particle physics in its gauge sector and/or particle sector is likely to give rise to additional (effective) interactions between Standard Model particles. In particular, in this work, we are concerned with such interactions experienced by the neutrinos when they propagate inside Earth matter. This effective term in the Lagrangian is of the form \cite{Wolfenstein:1977ue,Grossman:1995wx,Berezhiani:2001rs,Davidson:2003ha}
\begin{equation}
{\cal L}_{\rm NSI} = -2\sqrt{2}G_F\epsilon_{\alpha\beta}^{fC}(\overline{\nu_\alpha} \gamma^\mu P_L\nu_\beta)(\overline{f} \gamma_\mu P_C f) \,, 
\end{equation}
where $f$ is a fermion, $P_C=(1\pm\gamma_5)/2$ ($C = R,L$) are the chiral projection  operators, $G_F$ is the Fermi constant, and $\epsilon_{\alpha\beta}^{fC}$ are the corresponding NSI parameters. Since Earth matter is made up of the first generation fermions only, the NSI parameters corresponding to $e$, $u$, and $d$ are the only ones which contribute towards modifying the neutrino propagation inside the Earth. For the neutral-current NSI what is relevant is the sum $\epsilon^f_{\alpha\beta}= \epsilon_{\alpha\beta}^{fL} + \epsilon_{\alpha\beta} ^{fR}$. Furthermore, since only the incoherent sum of the NSI contributions is important, we combine the NSI effects coming from $\epsilon_{\alpha\beta}^{e}$, $\epsilon_{\alpha\beta}^{u}$, and $\epsilon_{\alpha\beta}^{d}$ as
\begin{equation}
\epsilon_{\alpha\beta} = \sum_{f=e,u,d}\frac{n_f}{n_e} \epsilon_{\alpha\beta}^{f} \,,
\end{equation}
where $n_f$ is the number density of the fermion $f$ and we have normalised the effective contribution to the number density of electrons in Earth.\footnote{Note that the convention followed in defining the NSI parameters is crucial to interpret the actual constraints on them from a given experiment and there are some places in the literature where a different convention has been followed 
(see Ref.~\cite{Ohlsson:2012kf} for a discussion).} 
While, in principle, the NSI parameters are complex, 
we consider only values which have phases either 0 and $\pi$. 
Throughout this work, we will use this assumption. Note that we sometimes refer to the NSI parameters as $\epsilon_{\alpha\beta}$ for simplicity, even though we work with only the real values of these parameters.

Each of the NSI parameters has been constrained from existing data. The corresponding model-independent upper bounds at 90~\% C.L.~are \cite{Biggio:2009nt}
\be
|\epsilon_{\alpha\beta}| < 
\begin{pmatrix}
4.2 & 0.33 & 3.0 \\ 0.33 & 0.068 & 0.33 \\ 3.0 & 0.33 & 21  
\end{pmatrix}
\,,
\label{eq:bound}
\ee
where we have arranged the parameters in the form of a matrix with the rows and columns corresponding to $\{ e,\mu,\tau\}$. Note that only the NSI parameter $\epsilon_{\mu\mu}$ is well constrained in this phenomenological approach, while constraints on all other NSI parameters are rather loose. In particular, large values of $\epsilon_{ee}$, $\epsilon_{e\tau}$, and $\epsilon_{\tau\tau}$ are still allowed. These bounds are generally referred to in the literature as indirect bounds as these bounds on parameters affecting neutrino oscillations come from non-neutrino experiments. The only neutrino experiment that has provided bounds on (some) of these parameters that are better than these indirect bounds is the SK experiment, which puts direct bounds on the NSI parameters $|\epsilon_{\mu\tau}|$ and $|\epsilon_{\tau\tau}-\epsilon_{\mu\tau}|$ in the framework of the so-called two-flavor hybrid model. From the combined SK~I and SK~II data sets, the 90~\% C.L.~upper bounds are given by \cite{Mitsuka:2011ty,Ohlsson:2012kf}
\be
|\epsilon_{\mu\tau}| < 0.033\,, \quad |\epsilon_{\tau\tau}-\epsilon_{\mu\mu}|< 0.147 
\,.
\label{eq:skbound}
\ee
The MINOS experiment has also used its data to set the following bound $-0.2 < \epsilon_{\mu\tau} < 0.07$ at 90~\% C.L.~\cite{Adamson:2013ovz}. However, this bound is weaker than the 
one set by SK.

In what follows, we will work with the exact three-flavor neutrino oscillation probabilities and consider the following ranges for the relevant NSI parameters:\footnote{We choose a smaller range 
than the current 90~\% C.L.~allowed range in Eq.~(\ref{eq:bound}), since outside these ranges 
the $\chi^2$ corresponding to the ICAL data is already large.}
\begin{eqnarray}
-0.3< & \epsilon_{e \mu} & < 0.3 \,, \nonumber\\
-0.5 <& \epsilon_{e\tau} & < 0.5 \,, \nonumber\\
-0.04 <& \epsilon_{\mu\tau} & < 0.04 \,, \nonumber\\ 
-0.15<& \epsilon_{\tau\tau} & < 0.15 \,, 
\label{eq:nsi}
\end{eqnarray}
while for the oscillation parameters we assume the following true values:
\be
\sin^2\theta_{12}=0.31 \,, \quad \sin^2\theta_{23}=0.5 \,, \quad \sin^22\theta_{13}=0.1 \,, \nn\\
\ms = 7.5\times 10^{-5}~{\rm eV}^2 \,, \quad |\ma|=2.4\times 10^{-3}~{\rm eV}^2 \,, \quad \delta_{\rm CP}=0 \,.
\label{eq:osc}
\ee

The NSI parameter(s), if present, will alter the neutrino oscillation probabilities. Oscillograms showing the impact of NSI on the neutrino oscillation probabilities have appeared vastly in the literature. The muon neutrino survival probability $P_{\mu\mu}$ is affected most by the parameters $|\epsilon_{\mu\tau}|$ and $|\epsilon_{\tau\tau}-\epsilon_{\mu\mu}|$, while the transition probability $P_{e\mu}$ depends on $|\epsilon_{e\mu}|$ and $|\epsilon_{\mu\tau}|$. This dependence can be used to discover  NSI parameters using neutrino oscillation data or constrain them. 

\begin{figure}[!t]
\begin{center}
\includegraphics[width=0.475\textwidth]{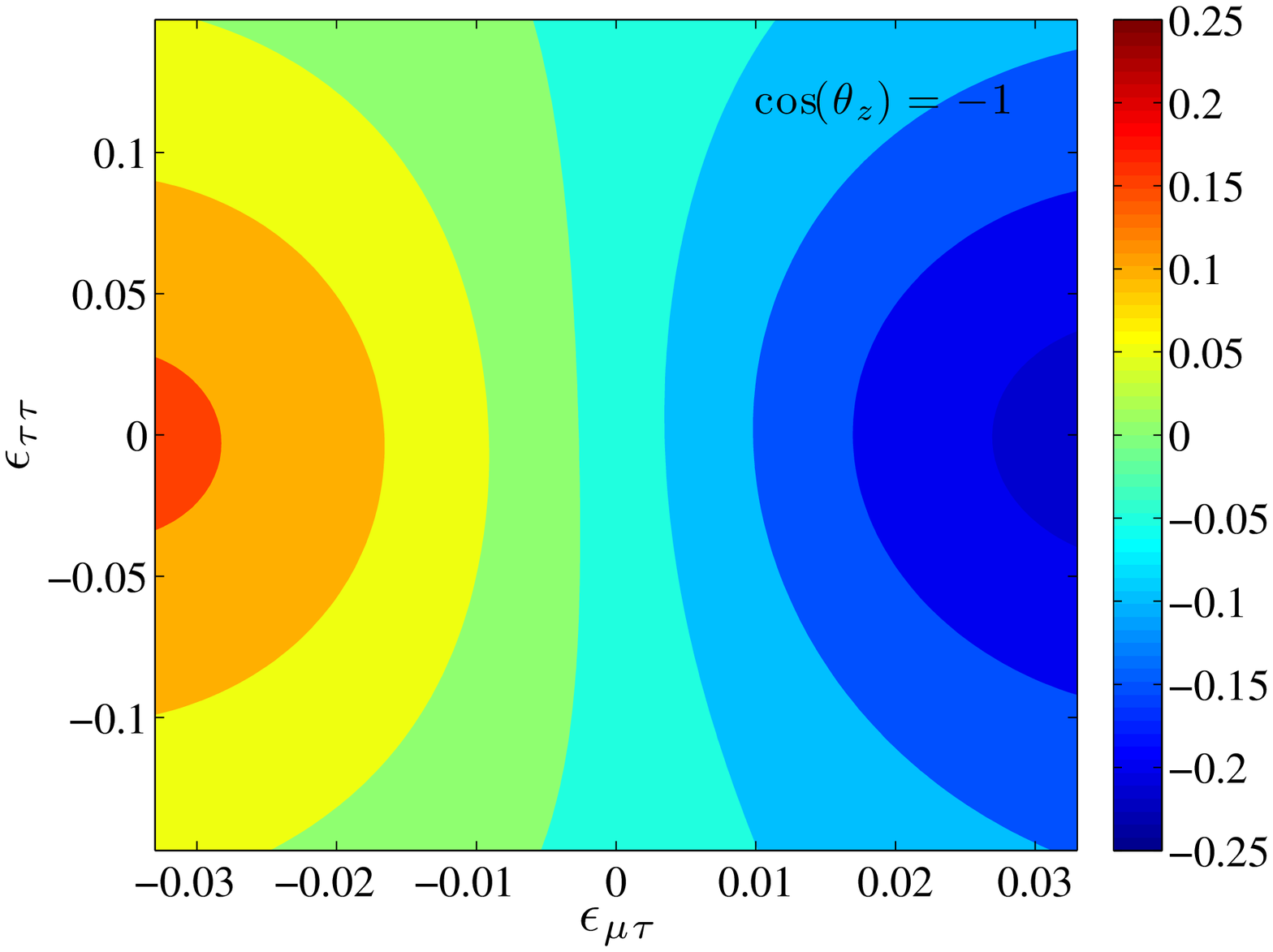}
\includegraphics[width=0.475\textwidth]{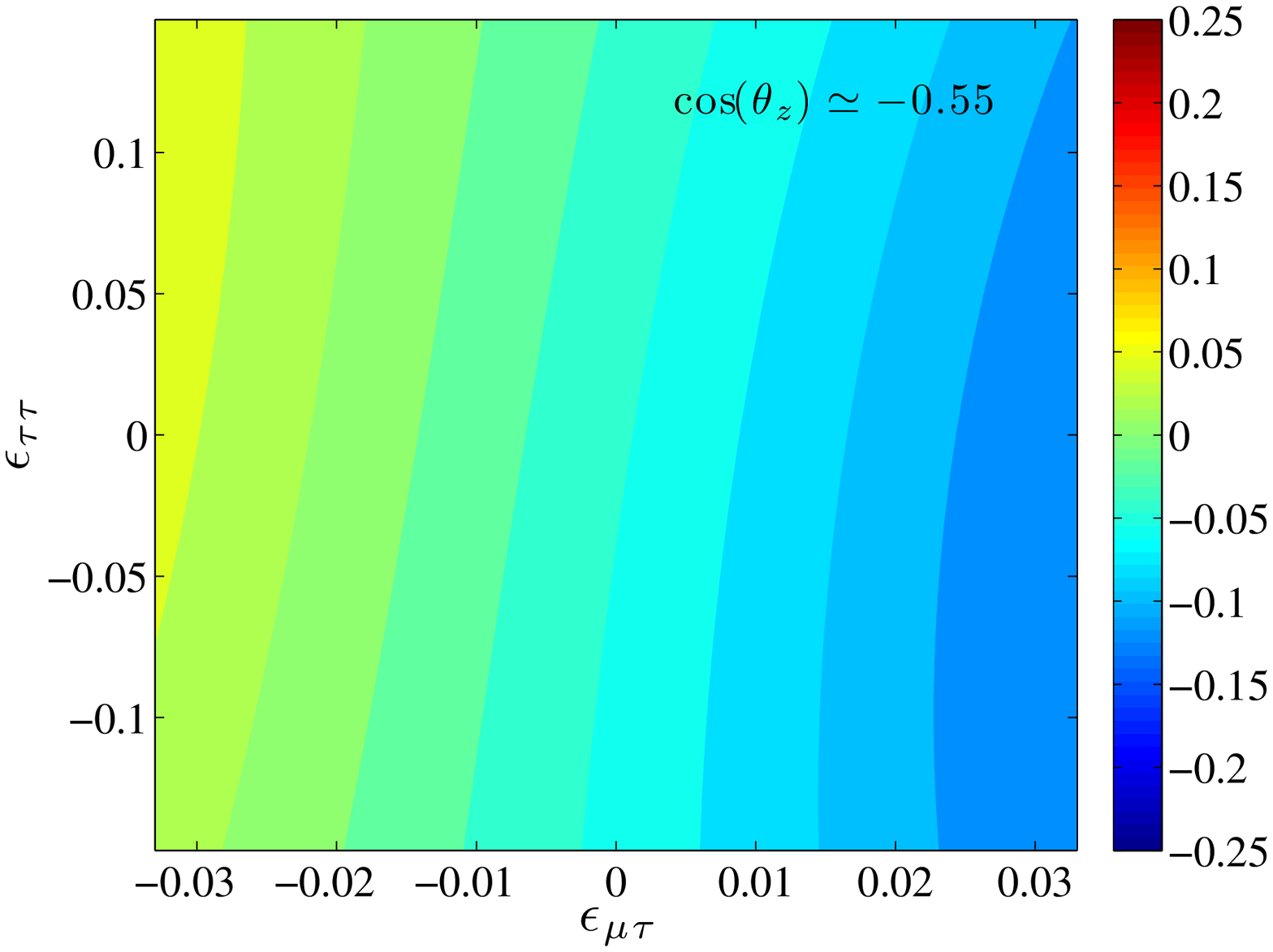}
\caption{The relative probability difference $A_{\mu\mu}^{\rm MH}$ as a function of the NSI parameters $\epsilon_{\mu\tau}$ and $\epsilon_{\tau\tau}$ for $\cos \theta = -1$ (left panel) and $\cos \theta = -0.55$ (right panel). Please note the scale of the colorbars to the right of each panel. The following values of the neutrino parameters have been used: $\theta_{12} = 34^\circ$, $\theta_{13} = 9.2^\circ$, $\theta_{23} = 45^\circ$, $\delta = 0$ (no leptonic CP-violation), $\Delta m_{21}^2 = 7.5 \cdot 10^{-5} \, {\rm eV}^2$, and $\Delta m_{31}^2 = + 2.4 \cdot 10^{-3} \, {\rm eV}^2$ (normal neutrino mass hierachy). All NSI parameters, except $\epsmt$ and $\epstt$, have been set to zero. 
}
\label{fig:pmm}
\end{center}
\end{figure}

\begin{figure}[!t]
\begin{center}
\includegraphics[width=0.475\textwidth]{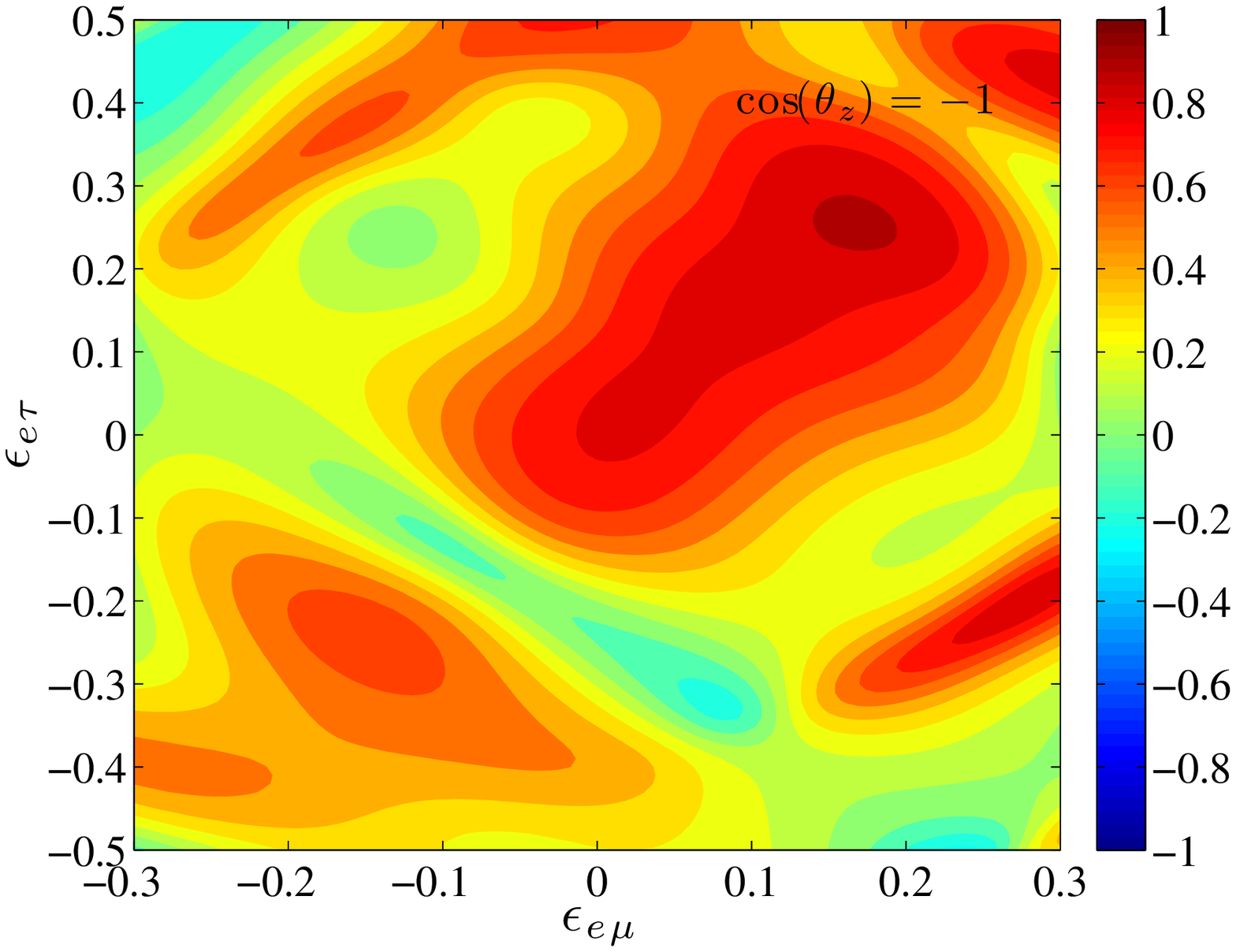}
\includegraphics[width=0.475\textwidth]{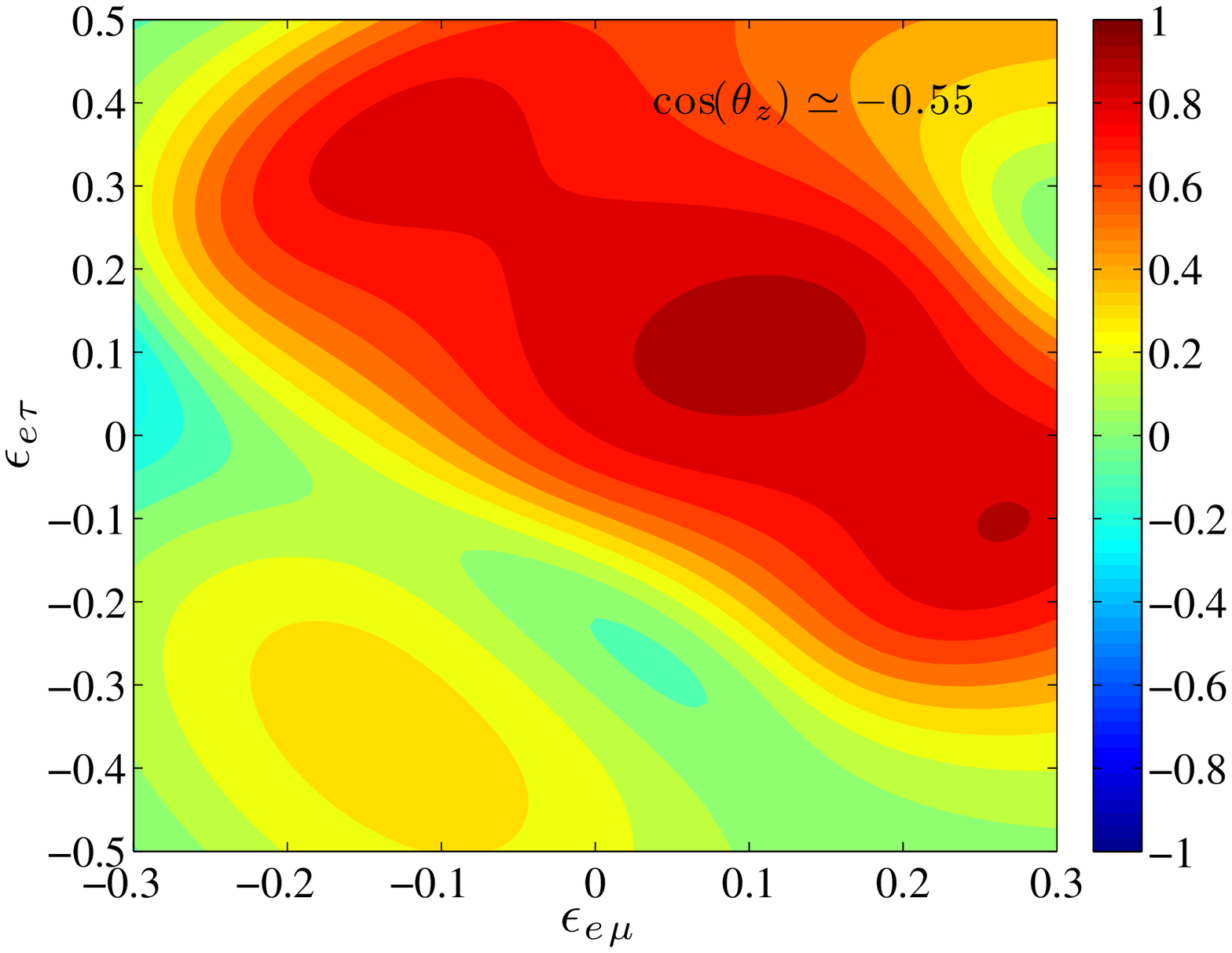}
\caption{The relative probability difference $A_{\mu e}^{\rm MH}$ as a function of the NSI parameters $\epsilon_{e\mu}$ and $\epsilon_{e\tau}$ for $\cos \theta = -1$ (left panel) and $\cos \theta = -0.55$ (right panel). The neutrino parameter values used are the same as in Fig.~\ref{fig:pmm}, except that $\epsem$ and $\epset$ are non-zero, while all other NSI parameter values have been set to zero. 
}
\label{fig:pem}
\end{center}
\end{figure}

It is also known that the dependence of the neutrino oscillation probabilities on NSI parameters is different for the NH and IH cases. 
Indeed, since measurement of the neutrino mass hierarchy is one of the prime goals of the INO experiment, it is pertinent to ask how the mass hierarchy sensitivity of the experiment alters in the presence of NSI. For the mass hierarchy determination, what matters is the difference in the oscillation probabilities between NH and IH. Therefore, it is appropriate to ask how the difference in the probabilities between NH and IH changes in presence of NSI. In order to show the impact of NSI on the mass hierarchy sensitivity, we present in Figs.~\ref{fig:pmm} and \ref{fig:pem} the contours of the relative difference $A_{\alpha\beta}^{\rm MH}$ between the neutrino oscillation probabilities $P_{\alpha\beta}$ (including NSI) corresponding to NH and IH. We define the relative probability difference $A_{\alpha\beta}^{\rm MH}$ as follows ({\emph{cf.}~the definition of the total CP-asymmetry in Ref.~\cite{Dick:1999ed})
\begin{equation}
A_{\alpha\beta}^{\rm MH}(\theta_z) = \frac{\Delta P_{\alpha\beta}^{\rm MH}(\theta_z)}{\Sigma P_{\alpha\beta}^{\rm MH}(\theta_z)}
=\frac{\int\Delta P_{\alpha\beta}^{\rm MH}(E,\theta_z) \, {\rm d}E}{\int\Sigma P_{\alpha\beta}^{\rm MH}(E,\theta_z) \, {\rm d}E} \,,
\end{equation}
where
\begin{eqnarray*}
\Delta P_{\alpha\beta}^{\rm MH}(E,\theta_z) &=& P_{\alpha\beta}^{\rm NH}(E,\theta_z) - P_{\alpha\beta}^{\rm IH}(E,\theta_z) \,,\nn \\
\Sigma P_{\alpha\beta}^{\rm MH}(E,\theta_z) &=& P_{\alpha\beta}^{\rm NH}(E,\theta_z) +P_{\alpha\beta}^{\rm IH}(E,\theta_z) \,,
\end{eqnarray*}
$P_{\alpha\beta}^{\rm NH}$ and $P_{\alpha\beta}^{\rm IH}$ being the $\nu_\alpha \to \nu_\beta$ oscillation probability for NH and IH, respectively. In each case, we calculate $A_{\alpha\beta}^{\rm MH}$ for a particular zenith angle $\theta_z$, while the energy dependence is integrated out in the range (1-100) GeV. 

In Fig.~\ref{fig:pmm}, we show the relative probability difference $A_{\mu\mu}^{\rm MH}$ in the $\epsilon_{\mu\tau}$-$\epsilon_{\tau\tau}$ plane, keeping $\epsilon_{e\mu}$ and $\epsilon_{e\tau}$ fixed at zero.\footnote{Here and throughout the rest of this work, we keep $\epsilon_{\mu\mu}=0$.} 
The probabilities are calculated numerically within the full three-generation oscillation 
paradigm, using the PREM \cite{Dziewonski:1981xy} density profile for the Earth matter. 
We compute this for two benchmark zenith angles of $\cos\theta_z = -1$ and $-0.55$ corresponding to neutrino baseline lengths of $L = 12742$~km and $L = 7000$~km, respectively. The colors represent the contours corresponding to the values of $A_{\mu\mu}^{\rm MH}$ shown in the colorbar. The (0,0) point in the $ \epsilon_{\mu\tau}$-$\epsilon_{\tau\tau}$ plane corresponds to neutrino oscillations without NSI (\ie~standard neutrino oscillations). 
At all other points, NSI are included in the model, and this can be 
observed to alter the mass hierarchy sensitivity parameter $A_{\mu\mu}^{\rm MH}$. 
Note that for standard oscillations we have $A_{\mu\mu}^{\rm MH} \sim -5\%$. This small 
relative difference is what the atmospheric neutrino experiments observing $\numu$ are 
exploiting to determine the neutrino mass hierarchy. When the NSI parameters are 
switched on, $A_{\mu\mu}^{\rm MH}$ changes. The relative difference $A_{\mu\mu}^{\rm MH}$
is seen to increase for $\epsilon_{\mu\tau} <0$ and decreases further to 
larger negative values for $\epsilon_{\mu\tau} > 0$. However, since for the hierarchy 
measurement what is relevant is the absolute difference $|A_{\mu\mu}^{\rm MH}|$, 
we will later see that for all $|\epsmt| >0$, the hierarchy sensitivity increases as long 
as all the parameters are kept fixed between the NH and IH cases. The 
change in $A_{\mu\mu}^{\rm MH}$ with $|\epsmt|$  is seen to be in the same direction for both the 
zenith angles shown, though its magnitude is seen to be larger for $\cos\theta_z=-1$ 
case.  The change in $A_{\mu\mu}^{\rm MH}$ with $\epstt$ is less pronounced. 
In particular, for the $\cos\theta_z=-0.55$ case, the dependence on $\epstt$ is marginal. 
Even for the $\cos\theta_z=-1$ case, the dependence of $A_{\mu\mu}^{\rm MH}$ on 
$\epstt$ for $\epsilon_{\tau\tau}=0$ is negligible. For larger values of $|\epsilon_{\tau\tau}|$, 
the role of $|\epstt|$ is to reduce the overall change in $A_{\mu\mu}^{\rm MH}$ due to NSI, 
and this happens for both positive and negative $\epstt$. 

In order to understand the change of the probability difference as a function of the NSI parameters, we can series expand the neutrino oscillation probabilities in orders of the NSI parameters and keep only the first-order terms. The expression for the difference
in the muon neutrino survival probability between NH and IH, 
keeping only leading-order terms in NSI parameters and neglecting the standard matter effects, is given by \cite{Kopp:2007ne,Ribeiro:2007ud,Kikuchi:2008vq}
\begin{eqnarray}
\Delta P_{\mu\mu}^{\rm MH} \simeq &&-2 {\rm Re}(\epsilon_{\mu\tau}) \sin 2\theta_{23} \bigg(\sin^22\theta_{23}
\frac{AL}{2E} \sin\frac{|\Delta m^2_{31}|L}{2E} + 4\cos^22\theta_{23}
\frac{A}{|\Delta m_{31}^2|} \sin^2\frac{|\Delta m^2_{31}|L}{4E}\bigg )\nn\\ &&
+(|\epsilon_{\mu\mu}| - |\epsilon_{\tau\tau}|  ) \sin^22\theta_{23}\cos2\theta_{23}
\bigg(\frac{AL}{2E}  \sin\frac{|\Delta m^2_{31}|L}{2E} -4\frac{A}{|\Delta m_{31}^2|} 
\sin^2\frac{|\Delta m^2_{31}|L}{4E}\bigg )
\,,
\label{eq:pmmapprox}
\end{eqnarray}
where $A = 2\sqrt{2} G_F n_e E$. 
Note that Eq.~(\ref{eq:pmmapprox}) depends only on the parameters ${\rm Re}(\epsilon_{\mu\tau})$ and $|\epsilon_{\mu\mu}| - |\epsilon_{\tau\tau}|$ to leading order. Dependence on $\epsilon_{e\mu}$ and $\epsilon_{e\tau}$ appear only at higher orders, which can be neglected unless these parameters are taken to be large. Therefore, in Fig.~\ref{fig:pmm}, we show the relative probability difference for the survival channel in the $\epsilon_{\mu\tau}$-$\epsilon_{\tau\tau}$ plane keeping the other NSI parameters at zero. For $\epsilon_{\tau\tau}=0$, the expression clearly shows that $|\Delta P_{\mu\mu}^{\rm MH} |$, and hence $|A_{\mu\mu}^{\rm MH} |$, grows with $|\epsilon_{\mu\tau}|$ and flips sign when the sign of $\epsilon_{\mu\tau}$ changes. The quantity $A_{\mu\mu}^{\rm MH}$ is positive for $\epsilon_{\mu\tau} <0$ and negative for $\epsilon_{\mu\tau}>0$. This agrees fairly well with the exact results shown in Fig.~\ref{fig:pmm}. The impact of $\epstt$ on the other hand is more involved. Using Eq.~(\ref{eq:pmmapprox}), we note that for any given large value of $\epsmt$, we should have the highest possible $|A_{\mu\mu}^{\rm MH}|$ for  $\epstt=0$, and since the dependence on this parameter comes in the form of $|\epstt|$, we should have lower $|A_{\mu\mu}^{\rm MH}|$ on both sides of $\epstt = 0$. On the other hand, for $\epsmt=0$, the 
$\Delta P_{\mu\mu}^{\rm MH}$ obtains contribution only from the second term,  
and there is a relative sign between the two terms in the parentheses. As a result for 
$\epsmt=0$ we do not expect large contribution to $\Delta P_{\mu\mu}^{\rm MH}$ from NSI. 
These features can be observed in the exact result in Fig.~\ref{fig:pmm}. 

In Fig.~\ref{fig:pem}, we present the $A_{e\mu}^{\rm MH}$ contours in the $\epsilon_{e\mu}$-$\epsilon_{e\tau}$ plane with $\epsilon_{\mu\tau}$ and $\epsilon_{\tau\tau}$ fixed at zero. The probability $P_{e\mu}$ depends crucially on the NSI parameters $\epsilon_{e\mu}$ and $\epsilon_{e\tau}$ at leading order, and hence, NSI bring significant change to $|A_{e\mu}^{\rm MH}|$. In this case, the corresponding analytic expression is complicated and we refer the reader to Ref.~\cite{Kopp:2007ne} for a related expression for the approximate formula. However, the exact results shown in the figure tell us that the presence of the NSI parameters $\epsilon_{e\mu}$ and $\epsilon_{e\tau}$ could bring substantial change to the mass hierarchy sensitivity of atmospheric neutrino experiments. In fact, $\epsem$ and $\epset$ could either increase or decrease the mass hierarchy sensitivity compared to what we expect from standard oscillations.

\section{\label{events}Event Spectrum at INO}

The ICAL (Iron CALorimeter) detector at INO will be a 50 kton 
detector with layers of magnetised iron interleaved with  
glass Resistive Plate Chambers (RPC), which will serve as the 
active detector element. The atmospheric neutrinos  
in $\nu_\mu$, $\overline{\nu_\mu}$, $\nu_e$, and $\overline{\nu_e}$ species come from 
decay of pions and kaons produced from cosmic ray interactions 
with the Earth's atmosphere. These neutrinos can interact with the  
detector nucleons producing the corresponding charged lepton. 
The dense iron material of ICAL helps to 
detect muons through their long tracks\footnote{The electrons give rise to an 
electromagnetic shower in the detector, which cannot travel far and is therefore 
more difficult to observe in this class of detectors.}, while the magnetic field 
allows the identification of their charge. Since Earth matter effects develop 
only in either the neutrino or the antineutrino channel for a given mass 
hierarchy, this charge identification 
capability gives ICAL an edge to better observe the Earth matter effects, 
and hence, the neutrino mass hierarchy. The capability of this 
experiment to help discover the mass hierarchy has been studied before 
by the INO collaboration using information on muon energy and angle  in Ref.~\cite{Ghosh:2012px} and 
using both the muon energy and angle information as well as hadron energy information in Ref.~\cite{Devi:2014yaa}. In this 
work, we only use the muon energy and angle information and quantify the change 
in the mass hierarchy sensitivity of ICAL in presence of NSI. We also 
study the prospects of constraining or discovering the NSI parameters 
with the muon event sample.

For calculating the predicted number of $\mu^-$ and $\mu^+$ events in ICAL, we use the 
same prescription as in the earlier INO collaboration papers.
The unoscillated events are calculated using the Nuance event generator 
modified for ICAL. 
The oscillation probabilities, with and without NSI effects,
are implemented through a re-weighting algorithm. Finally, the muon 
reconstruction efficiency, charge identification efficiency, and muon 
energy and angular resolutions are folded in as described in 
Refs.~\cite{Ghosh:2012px,Thakore:2013xqa,Ghosh:2013mga,Devi:2014yaa,Kaur:2014rxa}. 
The new ingredient in the simulations performed for this work is that while all 
the earlier papers used muon sample in the energy range (1-11)~GeV, we consider in this work a much larger energy range of (1-100)~GeV. 
In order to do that, we extend the earlier study \cite{Chatterjee:2014vta}
for muon detector response to 
100~GeV  from detector simulations done with the 
GEANT-based code developed for ICAL. The muon 
energy and zenith angle resolutions, as well as the charge 
identification efficiency and reconstruction efficiency  
are obtained as a function of muon energy and zenith angle. 
This is then folded with the oscillated events 
to obtain the final muon spectrum expected in ICAL. 
We generate raw events corresponding to 1000 years of running of 
ICAL in order to reduce the Monte Carlo fluctuations and normalise 
the final events to 10 years of running. 
This event sample is then 
binned in energy and zenith angle bins as follows. For the zenith angle 
we have 20 equal size bins in the $\cos\theta_z$ range ($-1$,1). For the energy, 
we take variable bin sizes to ensure that there are 
reasonable number of events in each bin. 
Between muon energies 
(1-10)~GeV, we take 9 energy bins of size 1~GeV, and between (10-100)~GeV, we 
take 3 energy bins of size 30~GeV. 

The predicted events are then analysed by a statistical procedure identical to 
the one used in the earlier papers. A $\chi^2$ function is defined as 
\be
\chi^2 = \chi^2(\mu^-) +  \chi^2(\mu^+)
\,,
\label{eq:chi1}
\ee
where 
\be
\chi^2(\mu^\pm) = \min_{\xi^\pm_k}\sum_{i=1}^{N_i}\sum_{j=1}^{N_j}
\bigg [ 2\bigg(N_{ij}^{\rm th}(\mu^\pm) - N_{ij}^{\rm ex}(\mu^\pm)\bigg ) + 
2N_{ij}^{\rm ex}(\mu^\pm)\ln\bigg(\frac{N_{ij}^{\rm ex}(\mu^\pm)}{N_{ij}^{\rm th}(\mu^\pm)}
\bigg ) \bigg]
+ \sum_{k=1}^l {\xi^\pm_k}^2
\,,
\label{eq:chi2}
\ee
\be
N_{ij}^{\rm th}(\mu^\pm) = {N'}_{ij}^{\rm th}(\mu^\pm)\bigg(1+\sum_{k=1}^l \pi_{ij}^k{\xi^\pm_k}\bigg)  
+{\cal O}({\xi^\pm_k}^2)\,,
\ee
${N'}_{ij}^{\rm th}(\mu^\pm)$ and $N_{ij}^{\rm ex}(\mu^\pm)$ being the predicted 
and `observed' number of $\mu^\pm$ events in ICAL, respectively, $\pi_{ij}^k$ 
the correction factors due to the $k^{th}$ systematic uncertainty, and $\xi^\pm_k$ the corresponding pull parameters. In this analysis, we include 5 systematic 
uncertainties. These are 20~\% error on flux normalisation, 10~\% error on 
cross-section, 5~\% uncorrelated error on the zenith angle distribution of atmospheric 
neutrino fluxes, 5~\% tilt error, and a 
5~\% overall error to account for detector systematics\footnote{Simulations to estimate the 
detector systematic uncertainties in ICAL is underway. This number could therefore 
change when better estimates of this become available.}. 
The individual contributions from 
$\mu^-$ and $\mu^+$ data samples are calculated by minimising Eq.~(\ref{eq:chi2})
over the pull parameters. These are then added [{\it cf.}~Eq.~(\ref{eq:chi2})] to obtain the 
$\chi^2$ for a given set of oscillation (and NSI) parameters. 
This resultant $\chi^2$ is then marginalised over the oscillation parameters, and when applicable, 
over the NSI parameters. We assume for the oscillation parameters the true values 
given in Eq.~(\ref{eq:osc}) and 
marginalise our $\chi^2$ over their current $3\sigma$ ranges. We include priors 
defined as 
\be
\chi^2_{\rm prior} = \bigg(\frac{p_{\rm true}-p}{\sigma_p}\bigg)^2\,,
\ee
where $p_{\rm true}$ is the assumed true value of the parameter $p$ and 
$\sigma_p$ the 1$\sigma$ error on it. We include priors on 
$|\Delta m_{31}^2|$, $\sin^2\theta_{23}$, and $\sin^22\theta_{13}$ with 
1$\sigma$ errors of 1~\%, 2~\%, and 0.005, respectively 
\cite{Agarwalla:2013qfa,Agarwalla:2014tpa}. For the 
NSI parameters, the $\chi^2$ is 
marginalised over their range given in Eq.~(\ref{eq:nsi}). 
For all results presented in this work, we use 500 kton-year of statistics in ICAL.
We next define the different 
$\chi^2$ that we compute in this work for the different physics studies we perform.

{\bf Sensitivity to Neutrino Mass Hierarchy:} To find the sensitivity of ICAL to the neutrino mass hierarchy, 
we compute the $\chi^2$ according to Eqs.~(\ref{eq:chi1}) and (\ref{eq:chi2}), where 
$N_{ij}^{\rm ex}(\mu^\pm)$ correspond to the right hierarchy and ${N'}_{ij}^{\rm th}(\mu^\pm)$ 
correspond to the wrong hierarchy. The 
$\Delta \chi^2_{\rm MH}$ thus obtained is then marginalised 
over the oscillation and NSI parameters. We do this for different assumed true values of the 
NSI parameters. 

{\bf Bounds on NSI parameters:} In the event that there is no signal for NSI in the 
ICAL data, one will be able to give upper bounds on the NSI parameters at a given C.L.
In order to find the expected sensitivity of ICAL 
to constrain the NSI parameters, we compute the $\chi^2$ by generating 
$N_{ij}^{\rm ex}(\mu^\pm)$ for standard oscillations (with all NSI parameters set to zero) 
and fitting this with ${N'}_{ij}^{\rm th}(\mu^\pm)$ computed with non-zero NSI parameters. 
The corresponding $\Delta \chi^2_{\rm S}$, where S stands for sensitivity, 
obtained after marginalising over oscillation gives a measure of the sensitivity reach of ICAL to NSI.

{\bf Discovery of NSI parameters:} If on the other hand, one finds a signal of NSI in the 
ICAL data, this would be a discovery of NSI, and hence, physics beyond the Standard 
Model. Of course, the NSI parameters in this case have to be above a certain value to be 
able to produce a discoverable signal at ICAL. We find this lower limit on the NSI 
parameters needed to be discovered at ICAL for a given C.L.~by generating 
$N_{ij}^{\rm ex}(\mu^\pm)$ with NSI and fitting this with ${N'}_{ij}^{\rm th}(\mu^\pm)$  
corresponding to standard oscillations. The corresponding $\Delta \chi^2_{\rm D}$ obtained 
after marginalising over oscillation  
gives a measure of the discovery reach of ICAL to NSI.

{\bf Precision on NSI parameters:} Finally, for a given set of NSI parameters, one 
can use the ICAL data to produce C.L.~contours in the NSI parameter space. We 
will show these contours at the 68~\%, 95~\%, and 99~\% C.L.~in the 
$\epsmt$-$\epstt$ and $\epsem$-$\epset$ planes. For this, we will generate 
$N_{ij}^{\rm ex}(\mu^\pm)$ for a certain set of NSI parameters and fit it with all values of the 
NSI parameters in the given plane while marginalising over the oscillation  
parameters. 

\section{\label{results}Expected Results from INO}

We now present our main results. We first show the impact of 
NSI on the mass hierarchy sensitivity of ICAL, which is the main goal of the 
experiment. We next give the sensitivity reach of this experiment in constraining 
NSI parameters. Subsequently, we quantify the NSI discovery potential at INO. 
Finally, we briefly discuss with what precision the NSI parameters could be 
measured at INO if they were indeed above the discovery limit. 

\subsection{\label{MH}Impact of NSI on Mass Hierarchy Sensitivity }

\begin{figure}[!t]
\begin{center}
\includegraphics[width=0.495\textwidth]{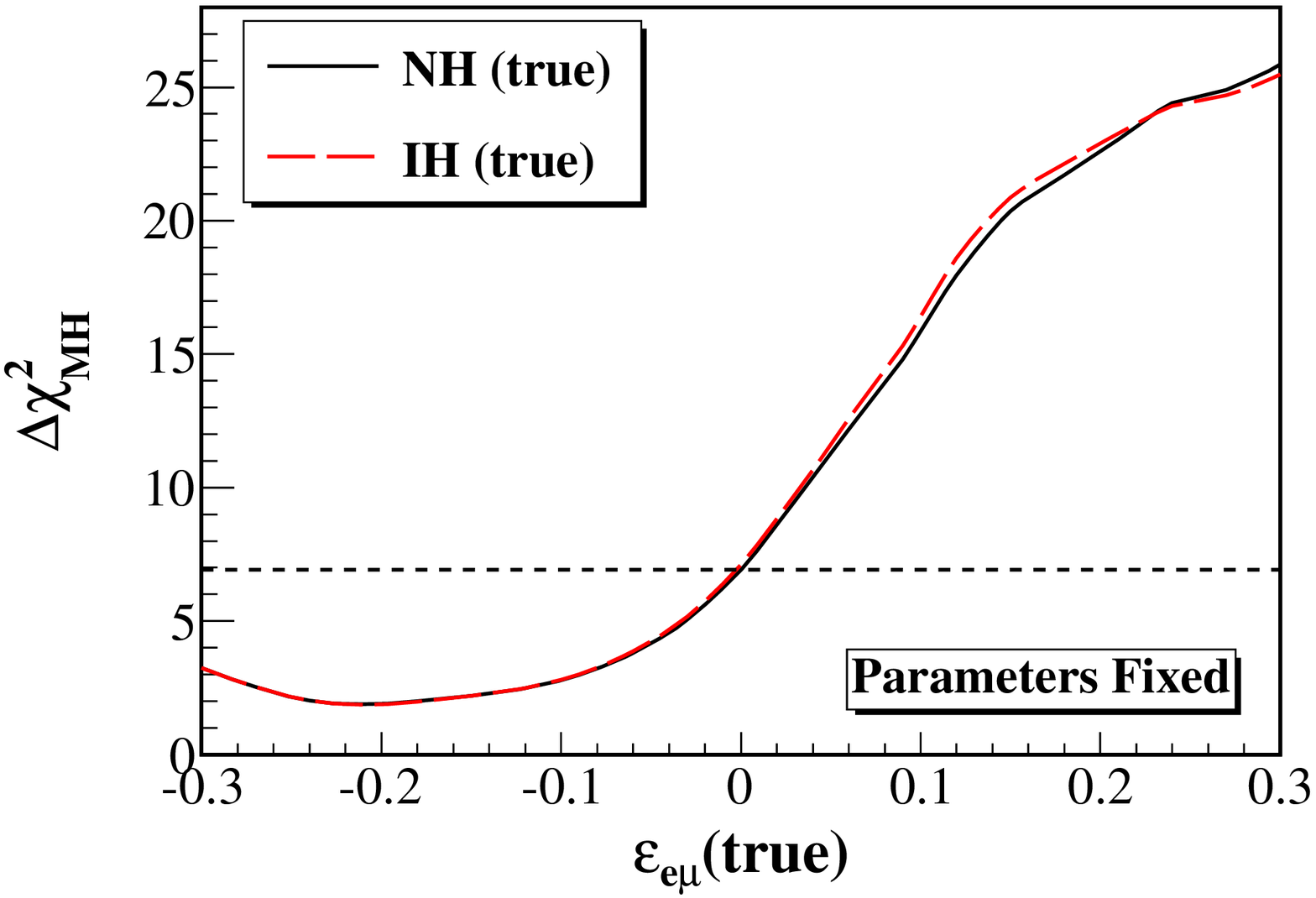}
\includegraphics[width=0.495\textwidth]{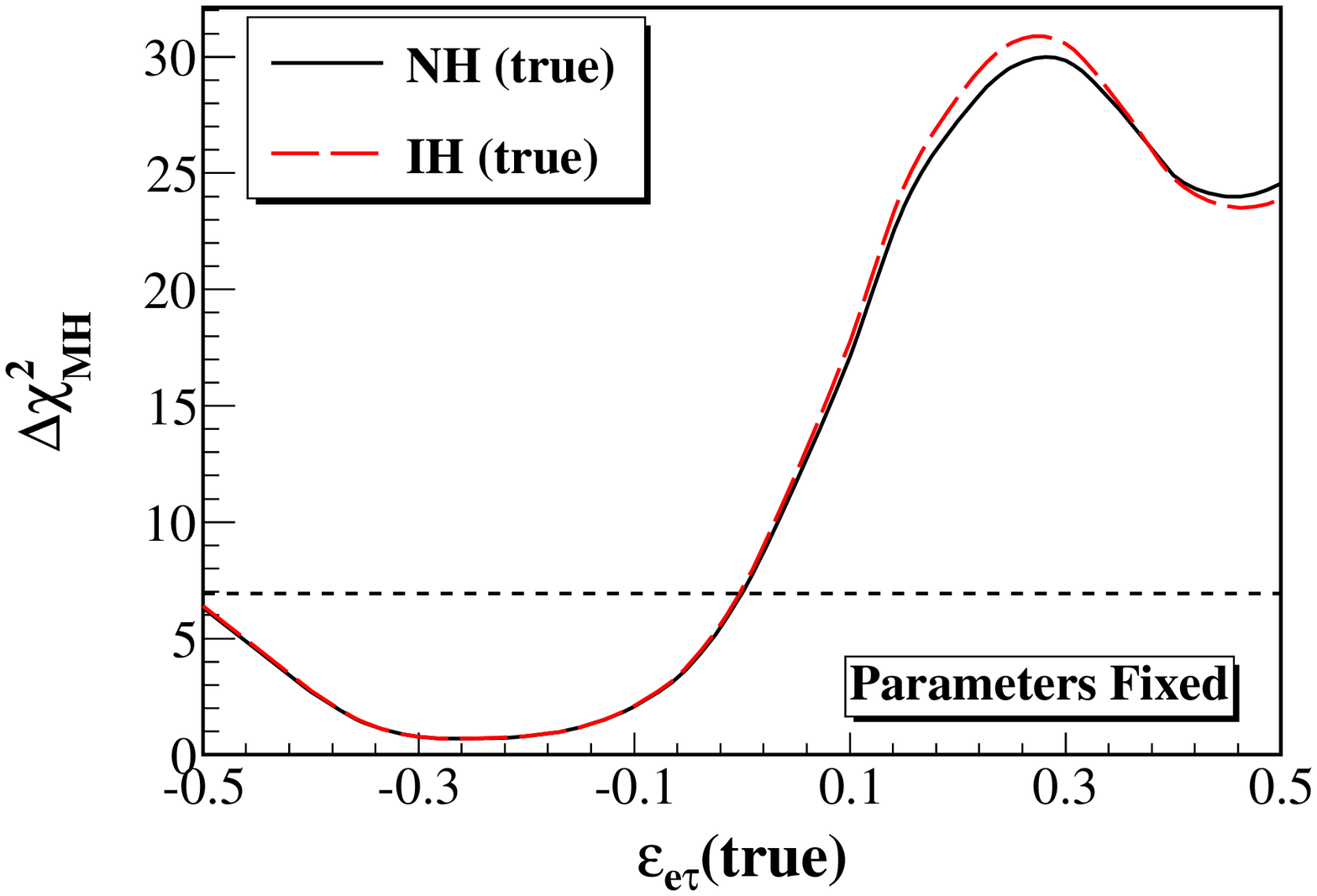}
\includegraphics[width=0.495\textwidth]{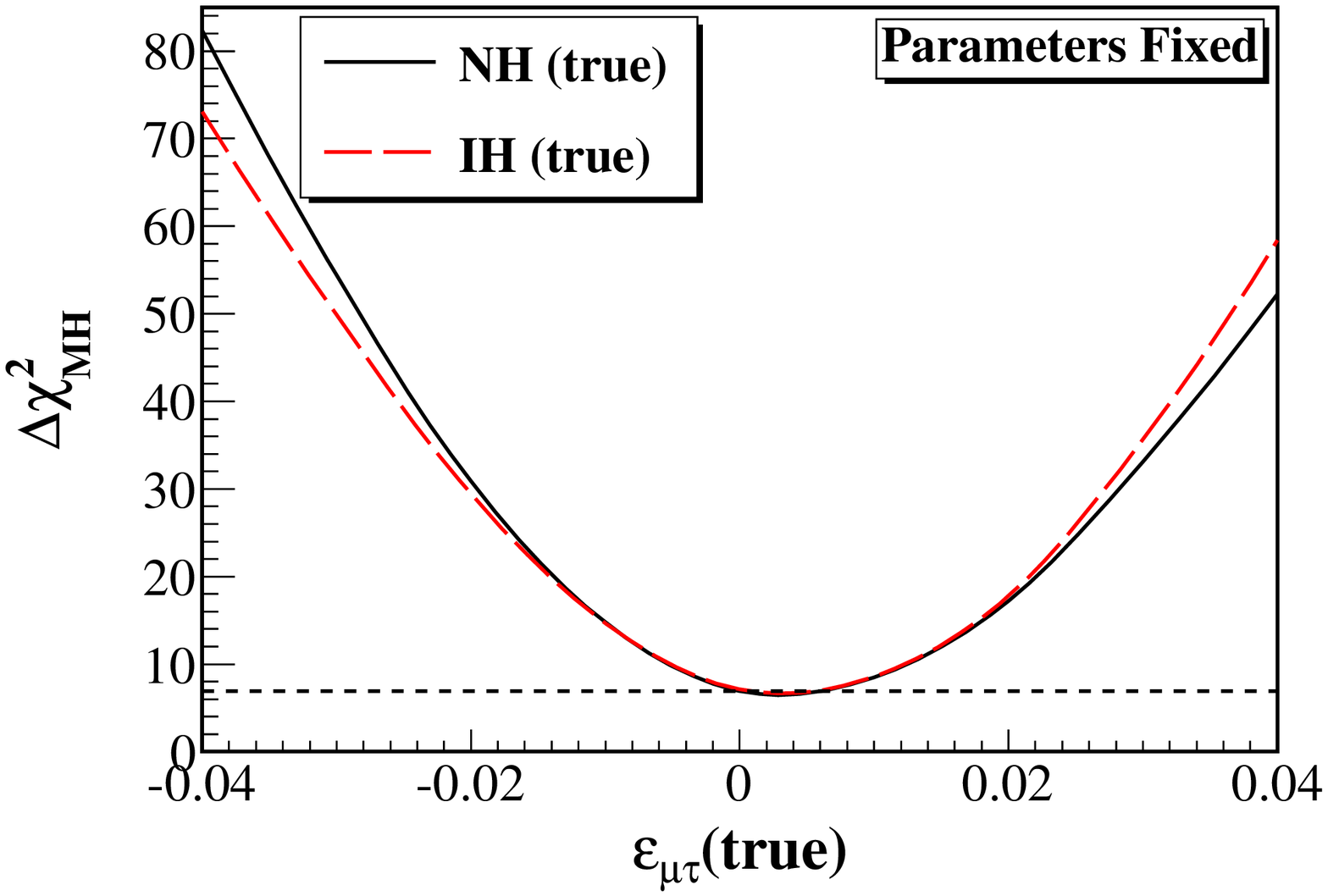}
\includegraphics[width=0.495\textwidth]{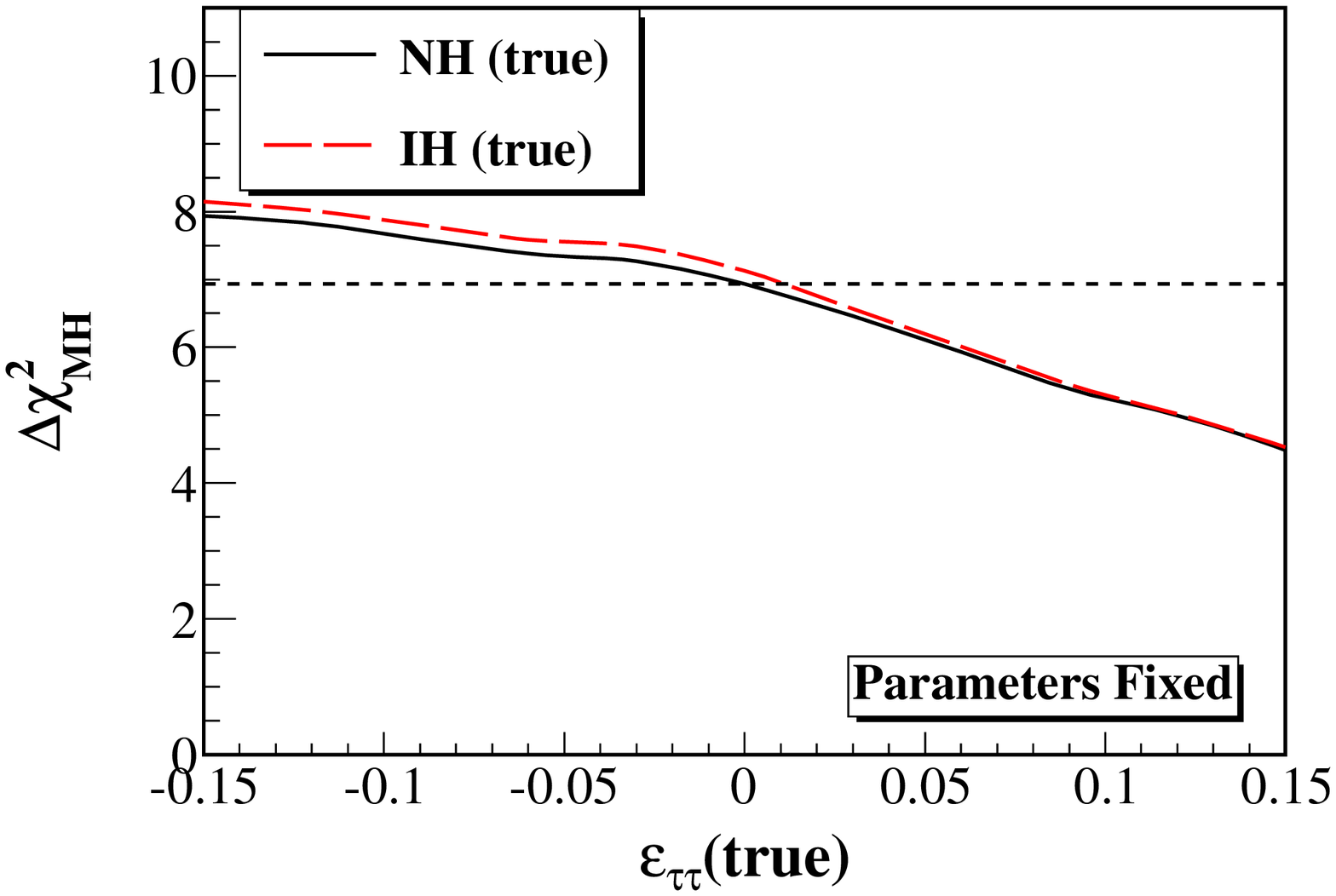}
\caption{The $\Delta \chi^2_{\rm MH}$, giving the 
expected mass hierarchy sensitivity from 10 years of running of ICAL, 
as a function of the true value of NSI parameters. We keep only one $\epsilon_{\alpha\beta}$(true)
to be non-zero at a time, while others are set to zero. The $\Delta \chi^2$ is 
obtained as explained in the text. However, the resultant $\Delta \chi^2$ is 
{\it not} marginalised over the oscillation parameters as well as NSI parameters. 
}
\label{fig:mhfixed}
\end{center}
\end{figure}

\begin{figure}[!t]
\begin{center}
\includegraphics[width=0.495\textwidth]{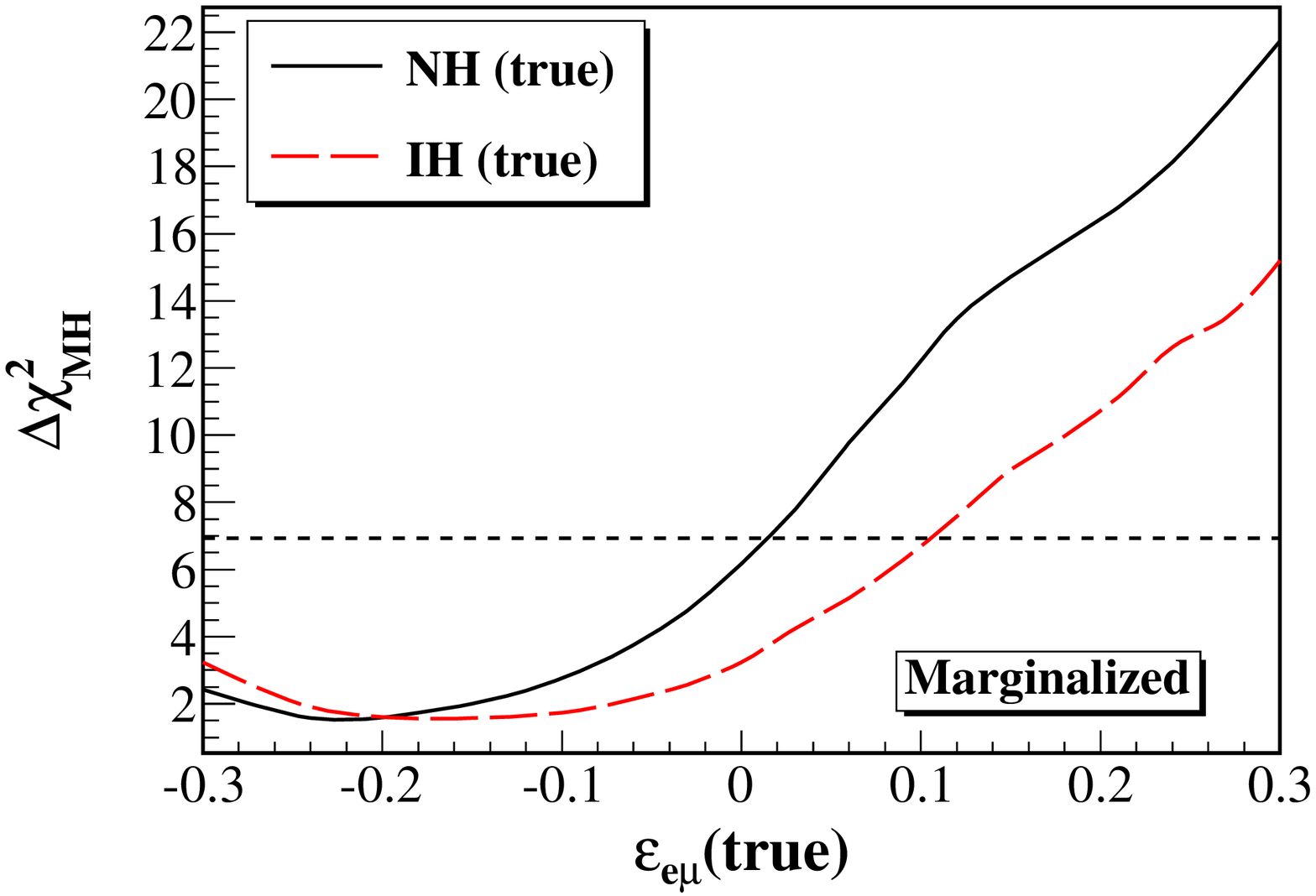}
\includegraphics[width=0.495\textwidth]{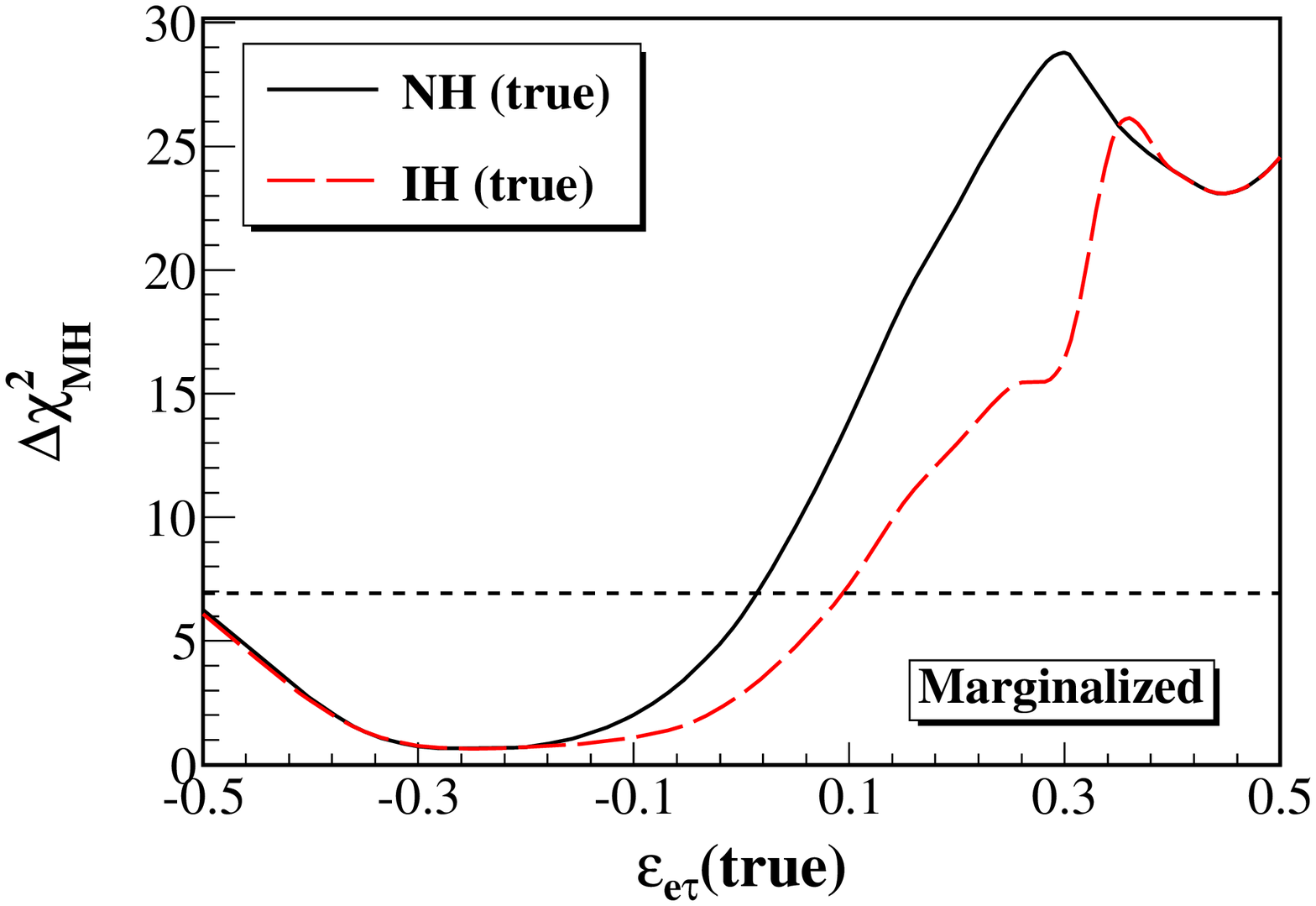}
\includegraphics[width=0.495\textwidth]{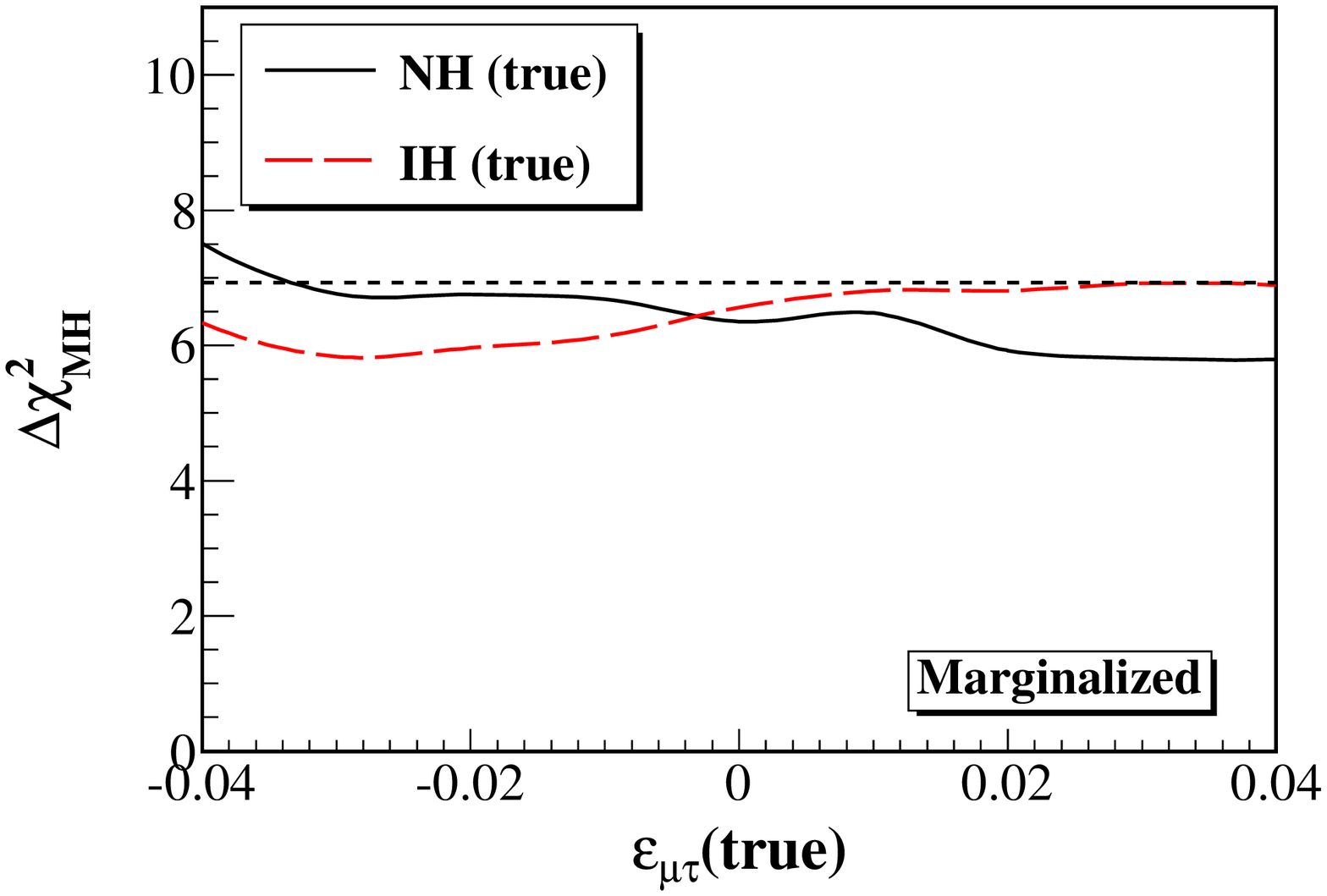}
\includegraphics[width=0.495\textwidth]{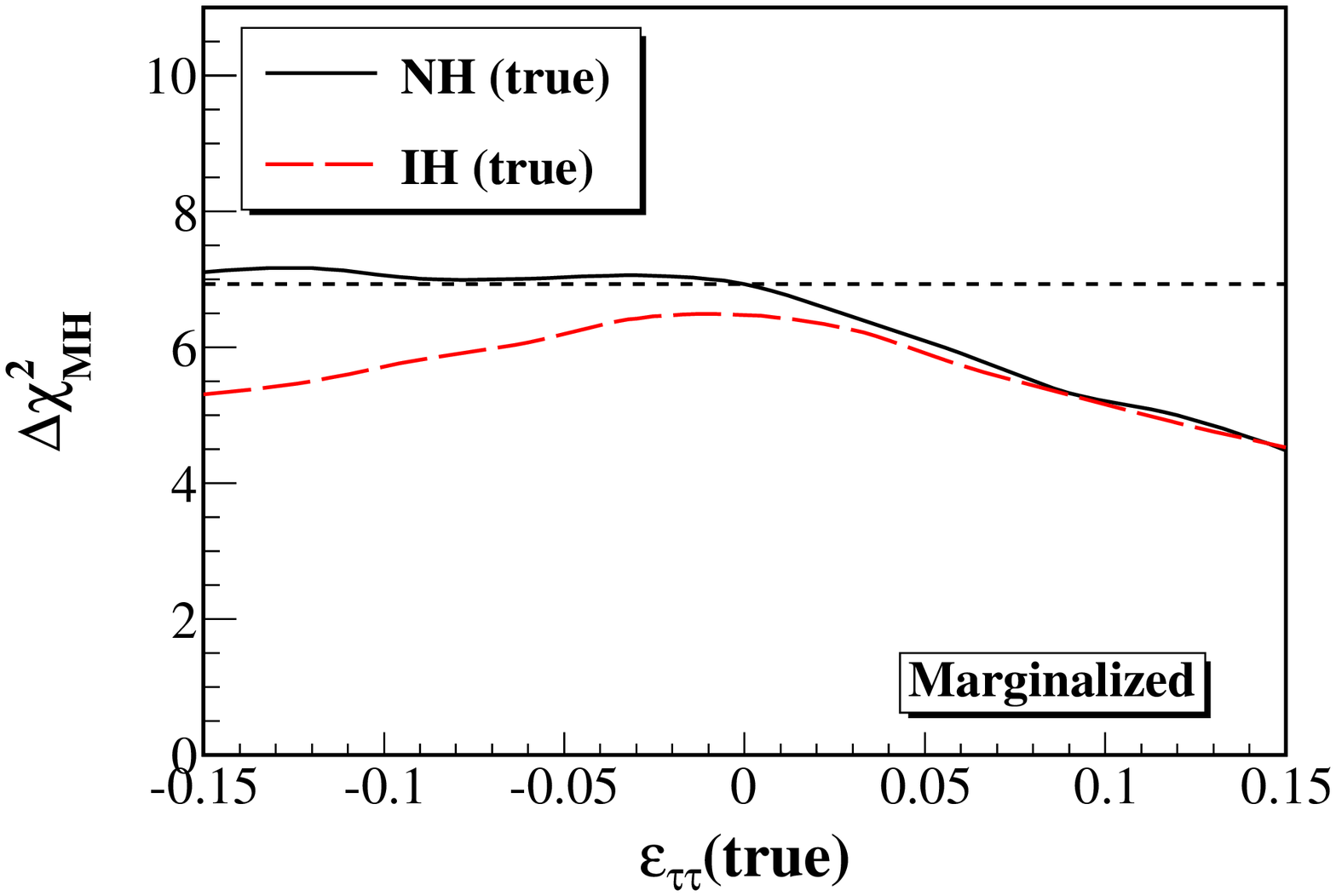}
\caption{The $\Delta \chi^2_{\rm MH}$, giving the 
expected mass hierarchy sensitivity from 10 years of running of ICAL, 
as a function of the true value of NSI parameters. 
We keep only one $\epsilon_{\alpha\beta}$(true)
to be non-zero at a time, while others are set to zero. 
The $\Delta \chi^2$ is 
obtained after marginalisation over the oscillation parameters as well as NSI parameters 
as explained in the text. 
}
\label{fig:mh}
\end{center}
\end{figure}

We noted in Figs.~\ref{fig:pmm} and \ref{fig:pem} that the difference in the 
oscillation probabilities between NH and IH changes in the presence of NSI.
Since the mass hierarchy sensitivity is defined in terms of the difference in the 
event distribution for NH and IH, it is therefore expected that 
the mass hierarchy sensitivity of the experiment would change 
in the presence of NSI. The mass hierarchy sensitivity for standard 
oscillations using only the muon events in ICAL is given in 
Ref.~\cite{Ghosh:2012px}. 
We revisit the mass hierarchy sensitivity in ICAL 
in the presence of NSI parameters and show our results in Figs.~\ref{fig:mhfixed} 
and \ref{fig:mh}. 
The $\Delta \chi^2_{\rm MH}$ corresponds to the difference in $\chi^2$ of the 
fit with the wrong and the right hierarchy as a function of the true value of 
the NSI parameter. For the sake of simplicity, we take only 
one non-zero NSI parameter in the data at a time. 
For instance, the black curves in the top-left panels of Figs.~\ref{fig:mhfixed} and \ref{fig:mh}
are obtained as follows. 
The data are generated 
for NH and a given true value of $\epsilon_{e\mu}$ (shown as the $x$-axis). 
The oscillation parameters in data are taken from Eq.~(\ref{eq:osc}) and 
all other NSI parameters are set to zero. 
This is then fitted with a theory prediction corresponding to  
IH. In Fig.~\ref{fig:mhfixed}, we present the $\Delta\chi^2_{\rm MH}$ 
obtained when all oscillation and NSI parameters in the fit are 
fixed at their assumed true values. In Fig.~\ref{fig:mh}, we 
marginalise the $\Delta\chi^2$ over the oscillation parameters 
$|\ma|$, $\sin^2\theta_{23}$, and $\sin^22\theta_{13}$ with priors.
The $\Delta\chi^2_{\rm MH}$ is also marginalised 
over the NSI parameter
which is taken as non-zero in the data, while the other NSI 
parameters are kept fixed at zero.  
For instance, in the top-left panel,
the $\Delta\chi^2$ is marginalised over $\epsilon_{e\mu}$,
while the other NSI parameters are kept fixed at zero. 
In all cases, the $\Delta\chi^2_{\rm MH}$ is marginalised over the oscillation 
parameters $|\ma|$, $\sin^2\theta_{23}$, and $\sin^2\theta_{13}$ with 
priors included as described in the previous section.
The other panels are also obtained in a similar way. 

The horizontal black 
dashed lines in the four panels of Figs.~\ref{fig:mhfixed} and \ref{fig:mh}
show the mass hierarchy sensitivity 
expected in ICAL for the case when there are no NSI considered in either 
the data or the fit. A comparison of this with the black solid and red dashed curves 
in the figure reveals that presence of NSI in the data could change the mass 
hierarchy sensitivity of ICAL. In particular, we see that the $\Delta \chi^2_{\rm MH}$ 
changes sharply with the true value of $\epsilon_{e\mu}$ and 
$\epsilon_{e\tau}$. In presence of NSI, we note that the $\Delta \chi^2_{\rm MH}$ 
increases 
for $\epsilon_{e\mu}({\rm true})\gtrsim 0$ and $\epsilon_{e\tau}({\rm true})\gtrsim 0$, 
while it decreases for 
$\epsilon_{e\mu}({\rm true})\lesssim 0$ and $\epsilon_{e\tau}({\rm true})\lesssim 0$, 
compared to what is expected for standard oscillations.  

These features can be understood from Figs.~\ref{fig:pmm} and 
\ref{fig:pem}.  Since the NSI parameters $\epsilon_{e\mu}$ and $\epsilon_{e\tau}$ 
mainly affect the appearance channel $P_{e\mu}$, we refer to Fig.~\ref{fig:pem} to 
understand the upper panels of Figs.~\ref{fig:mhfixed} and \ref{fig:mh}. 
The (0,0) point of Fig.~\ref{fig:pem} 
refers to standard oscillations and gives the mass hierarchy sensitivity shown by the 
black dashed lines in Figs.~\ref{fig:mhfixed} and \ref{fig:mh}. If we 
stay on $\epsilon_{e\tau}=0$ and 
change $\epsilon_{e\mu}$, we note from the left panel ($\cos\theta_z=-0.55$) 
of Fig.~\ref{fig:pem} 
that for $\epsilon_{e\mu}\lesssim 0$  $|A_{e\mu}^{\rm MH}|$ decreases, while for 
$\epsilon_{e\mu}\gtrsim 0$ it increases. This is less clear in 
the core-crossing bin, however, 
since the largest mass hierarchy sensitivity at ICAL comes from zenith angle bins 
close to $\cos\theta_z=-0.55$, this feature stays in the final $\Delta \chi^2_{\rm MH}$.

The effect of $\epsilon_{\tau\tau}$ on the mass hierarchy sensitivity 
is seen to be less severe from the bottom-right panels of 
Figs.~\ref{fig:mhfixed} and \ref{fig:mh}. This NSI parameter affects the muon 
neutrino survival channel the most. Figure~\ref{fig:pmm} reveals that the impact of 
$\epsilon_{\tau\tau}$ on $|A_{\mu\mu}^{\rm MH}|$ (when $\epsilon_{\mu\tau}=0$)
is very small for both the core-crossing and the $\cos\theta_z=-0.55$ bin.  
The impact of $\epsilon_{\mu\tau}$ on the mass hierarchy sensitivity is more 
interesting and has been discussed in 
Refs.~\cite{Choubey:2014iia,Mocioiu:2014gua,Chatterjee:2014gxa}. 
For $\epsilon_{\tau\tau}=0$, we see that  $|A_{\mu\mu}^{\rm MH}|$ could change 
up to 20~\% for the core-crossing bin and a few percent for the $\cos\theta_z=-0.55$ bin, as we change $\epsilon_{\mu\tau}$.  
Note that for standard oscillation $|A_{\mu\mu}^{\rm MH}|$ is already a very small 
number, and hence, the relative change of $|A_{\mu\mu}^{\rm MH}|$ due to 
$\epsilon_{\mu\tau}$ is significant. This is reflected in the bottom-left 
panel of Fig.~\ref{fig:mhfixed}, where we see a large increase in $\Delta\chi^2_{\rm MH}$ 
with $\epsilon_{\mu\tau}$. However, once we marginalise over oscillation and 
$\epsilon_{\mu\tau}$ in the fit, this increase is washed out and we obtain no 
significant impact of $\epsilon_{\mu\tau}$ on $\Delta\chi^2_{\rm MH}$ in the bottom-left panel of Fig.~\ref{fig:mh}.

\subsection{\label{bounds}Expected Bounds on NSI}

\begin{figure}[!t]
\begin{center}
\includegraphics[width=0.495\textwidth]{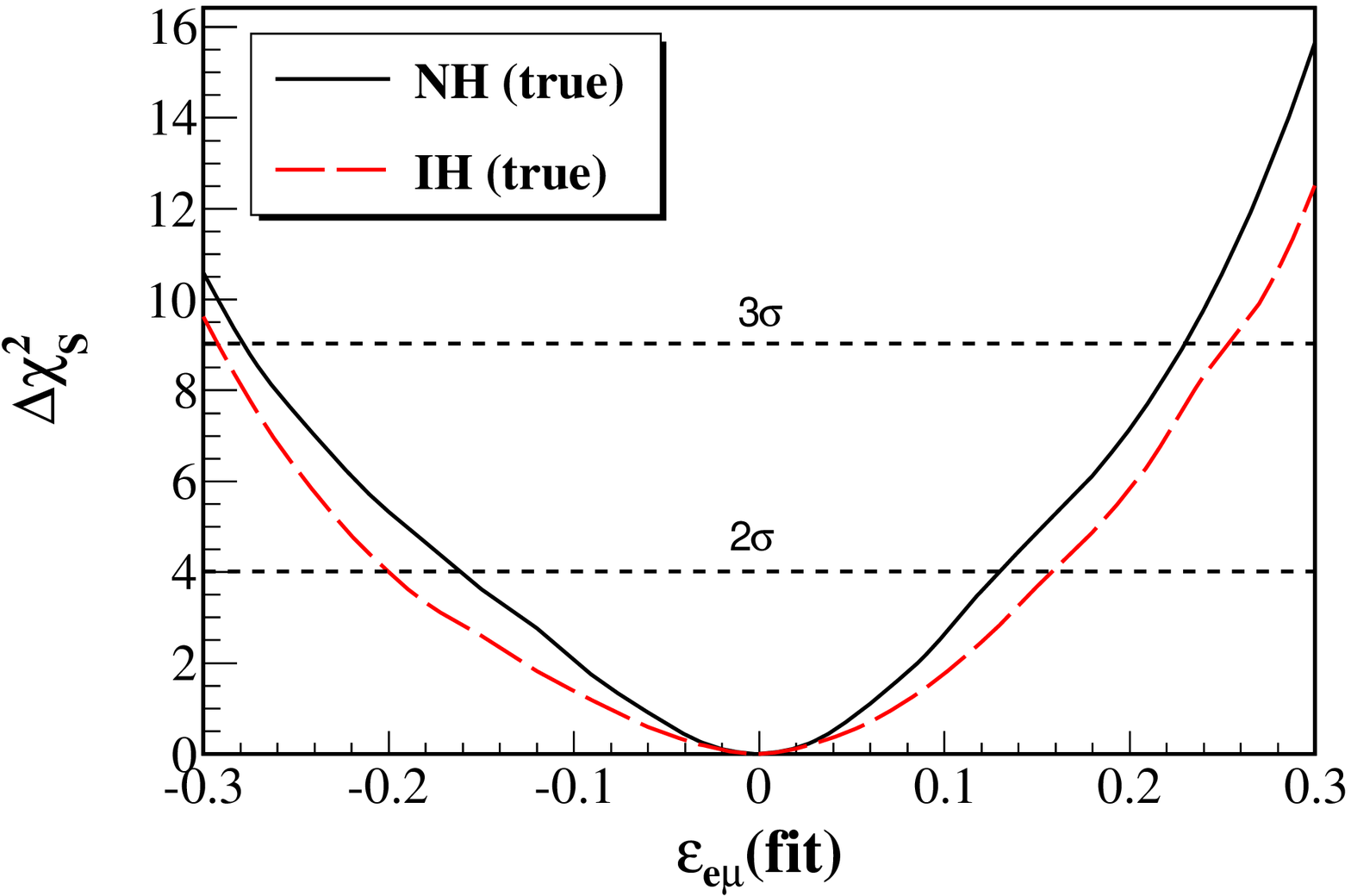}
\includegraphics[width=0.495\textwidth]{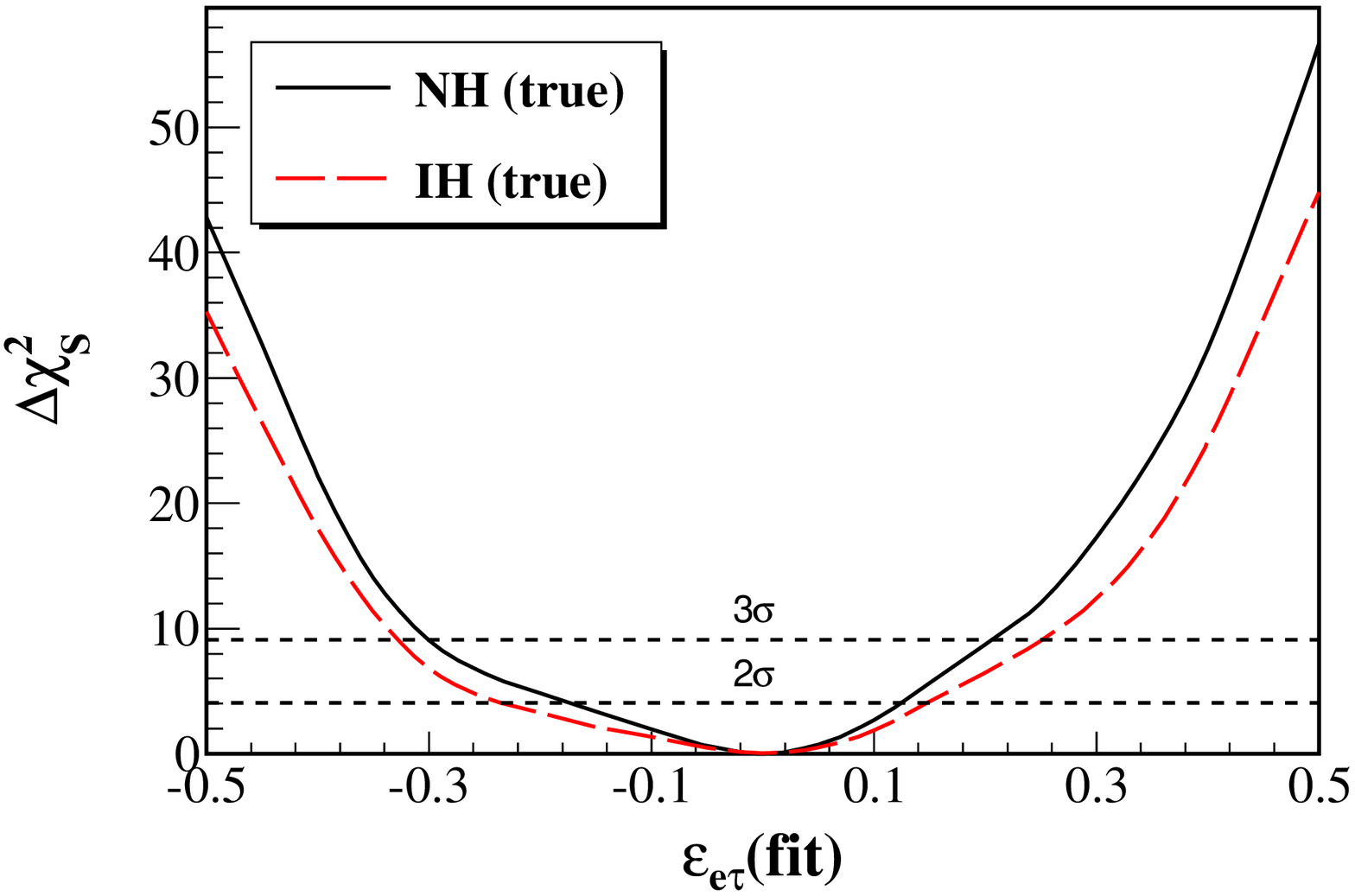}
\includegraphics[width=0.495\textwidth]{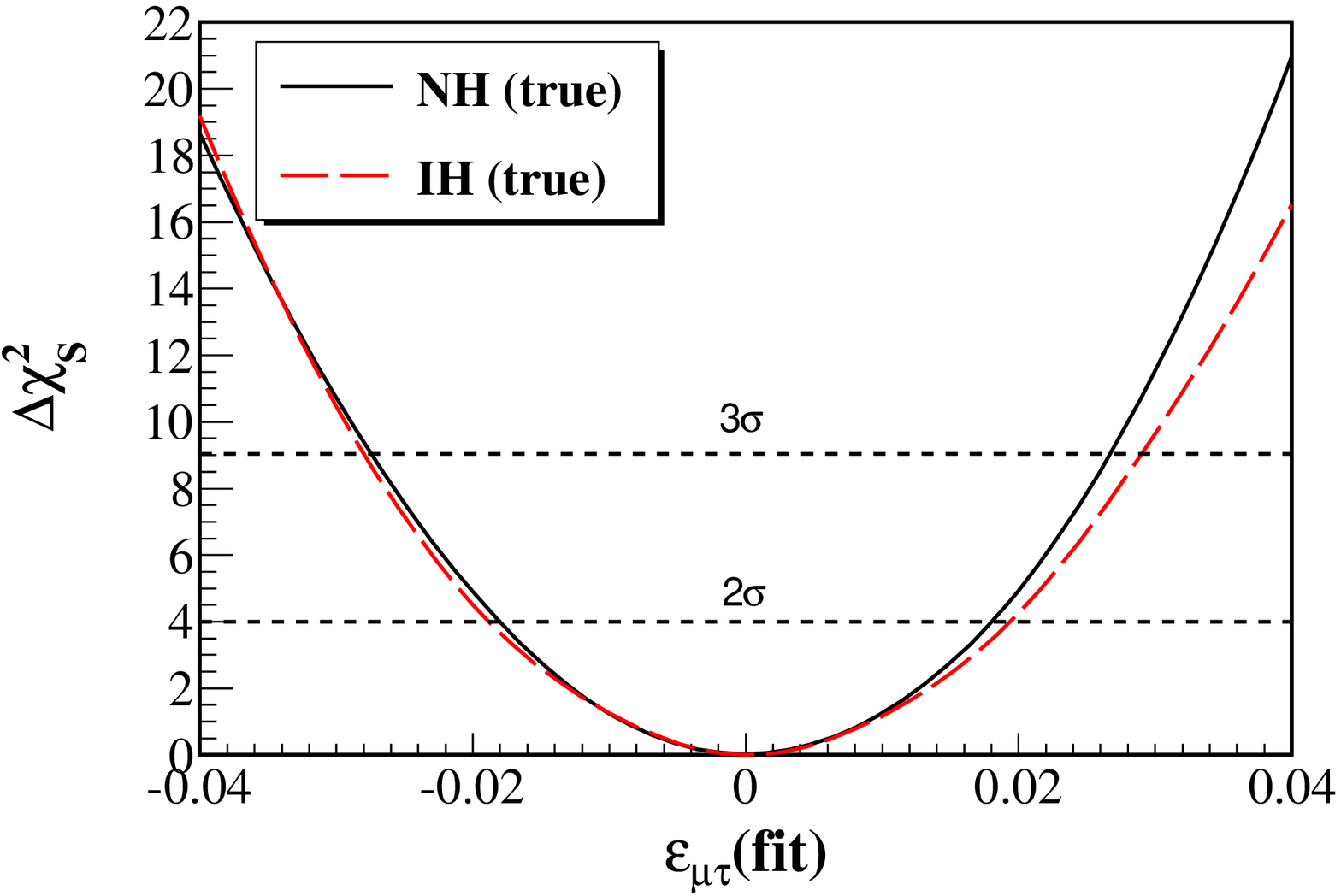}
\includegraphics[width=0.495\textwidth]{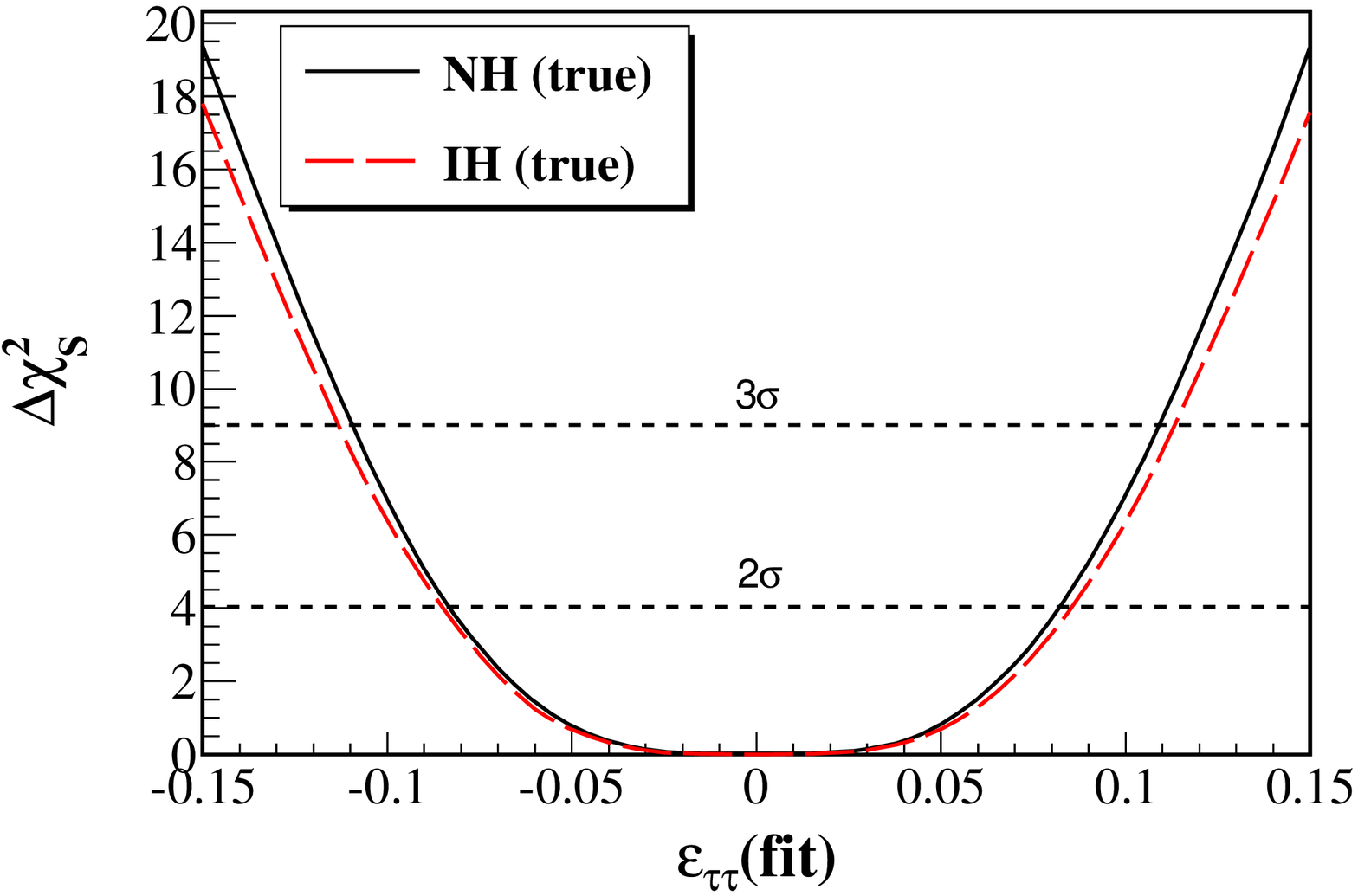}
\caption{The $\Delta \chi^2_{\rm S}$, giving the sensitivity reach of 10 years of 
ICAL data in constraining the NSI parameters in the event that the data 
show no signal of any new physics, as a function of the fit value of NSI parameters. 
We keep only one $\epsilon_{\alpha\beta}$(true) 
to be non-zero at a time, while others are set to zero. 
The $\Delta \chi^2_{\rm S}$ has been 
marginalised over the oscillation parameters and NSI parameters as explained 
in the text.
}
\label{fig:sens}
\end{center}
\end{figure}

In Fig.~\ref{fig:sens}, we present the expected sensitivity reach of ICAL 
in constraining NSI parameters. 
The figure shows the 
expected sensitivity for each of the NSI parameters 
$\epsilon_{e\mu}$ (top-left panel),
$\epsilon_{e\tau}$ (top-right panel),
$\epsilon_{\mu\tau}$ (bottom-left panel), and 
$\epsilon_{\tau\tau}$ (bottom-right panel). 
This figure is obtained as follows. We use  
as data the event distribution at ICAL corresponding to standard oscillations 
by setting all NSI parameters to zero. This is then fitted with 
the predicted event distribution which includes one non-zero NSI parameter at a time, and the corresponding 
$\Delta \chi^2_{\rm S}$ calculated. In Fig~\ref{fig:sens}, we show this $\Delta \chi^2_{\rm S}$ as a 
function of the NSI parameter that is allowed to be non-zero in the fit. The $\Delta \chi^2_{\rm S}$ is marginalised over 
the oscillation parameters 
 $|\ma|$, $\sin^2\theta_{23}$, and $\sin^22\theta_{13}$. 
 Priors on the three oscillation parameters were included as described in the previous section. 
 The resultant 
 $\Delta \chi^2_{\rm S}$ shows little change as a result of marginalisation over them. 
 The black solid curves are obtained when the data are considered corresponding to 
NH, while the red dashed curves are for data corresponding to IH. We keep the 
hierarchy fixed to its assumed true value in the fit. 

The expected sensitivity for IH is only marginally worse than that for NH. 
At the 90~\% (3$\sigma$) C.L., the expected bounds on the NSI parameters 
from 500 kton-years of statistics in ICAL for NH can be read from the figure as
\be
-0.119~(-0.3)& <  \epsem <  &0.102~(0.2) \,, \nn \\
-0.127~(-0.27) &<  \epset <  &0.1~(0.23) \,, \nn \\
-0.015~(-0.027) &<  \epsmt <  &0.015~(0.027) \,, \nn \\
-0.073~(-0.109)& <  \epstt <  &0.073~(0.109) \,. \nn 
\ee
For the IH case, the bounds are comparable and can be read from the figure.

\begin{figure}[!t]
\begin{center}
\includegraphics[width=0.495\textwidth]{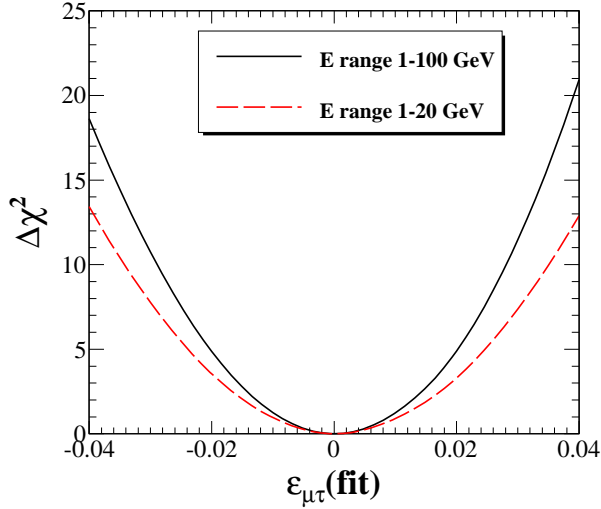}
\caption{The improvement in the expected bounds on NSI 
from increasing the considered muon energy range from 20~GeV to 100~GeV 
in the analysis. We show this only for the NSI parameters $\epsmt$. 
}
\label{fig:compare}
\end{center}
\end{figure}

In Fig.~\ref{fig:compare}, we show the improvement that we obtain in the sensitivity 
reach of ICAL to the NSI parameters when we increase the muon energy 
range considered in the analysis from 20 GeV (red dashed curve) to 
100 GeV (black solid curve). The 3$\sigma$ bound on $\epsmt$ improves from 
$-0.033 < \epsmt < 0.033$ to $-0.027 <  \epsmt <  0.027$, when we increase the muon energy 
from 20 GeV to 100 GeV in the data. 
 
\subsection{\label{discovery}Discovery Reach for NSI Parameters}

\begin{figure}[!t]
\begin{center}
\includegraphics[width=0.495\textwidth]{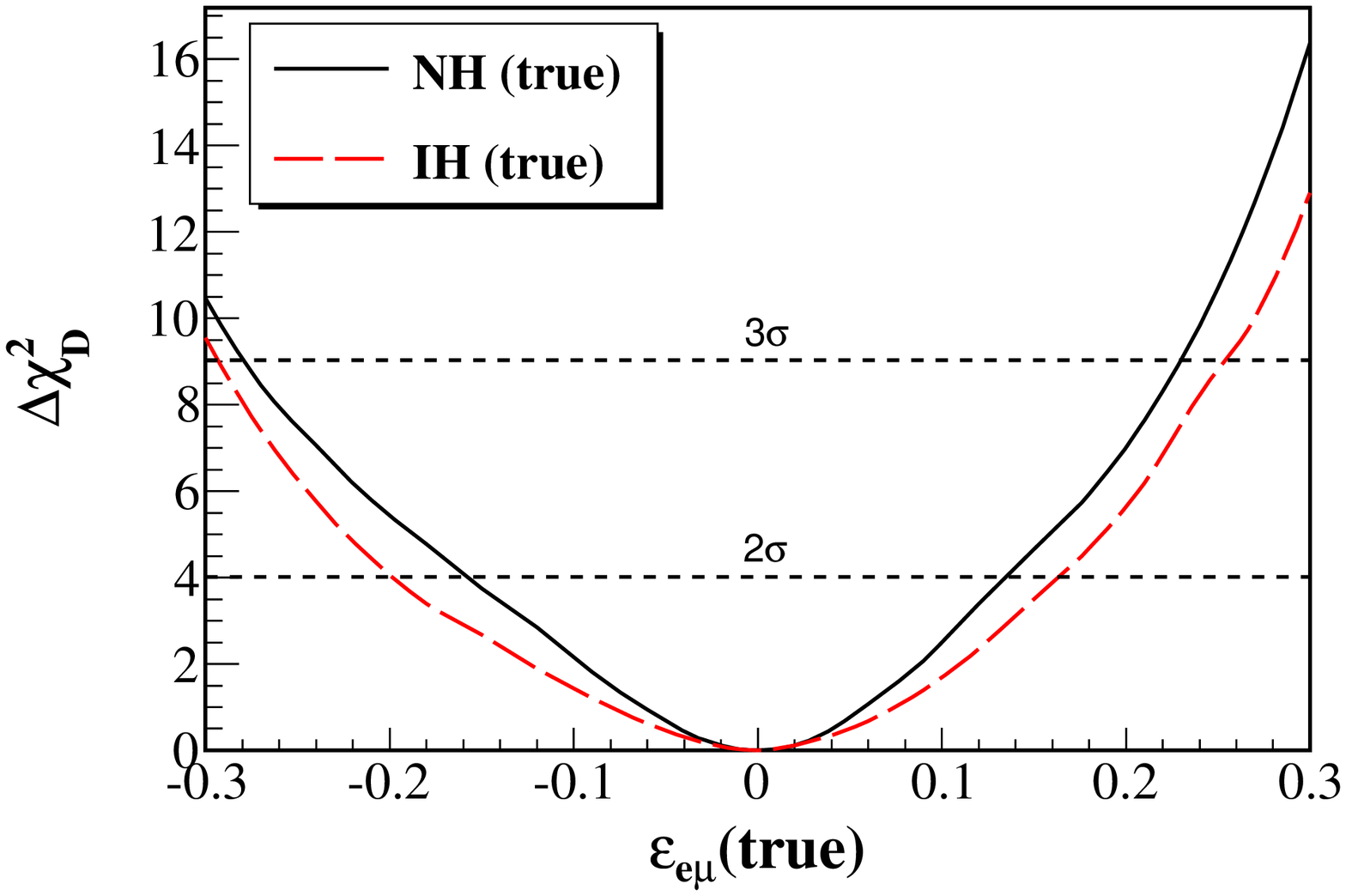}
\includegraphics[width=0.495\textwidth]{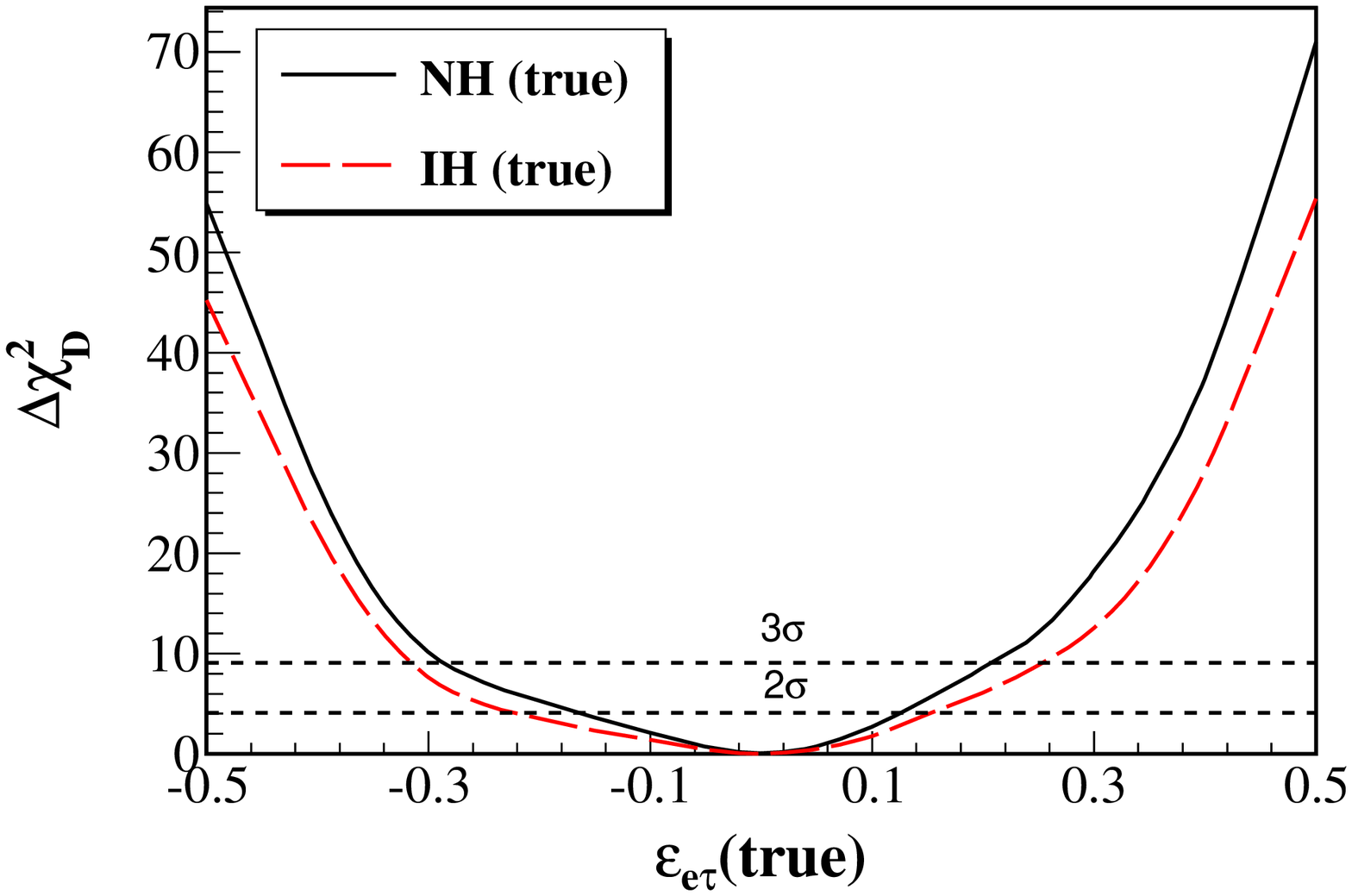}
\includegraphics[width=0.495\textwidth]{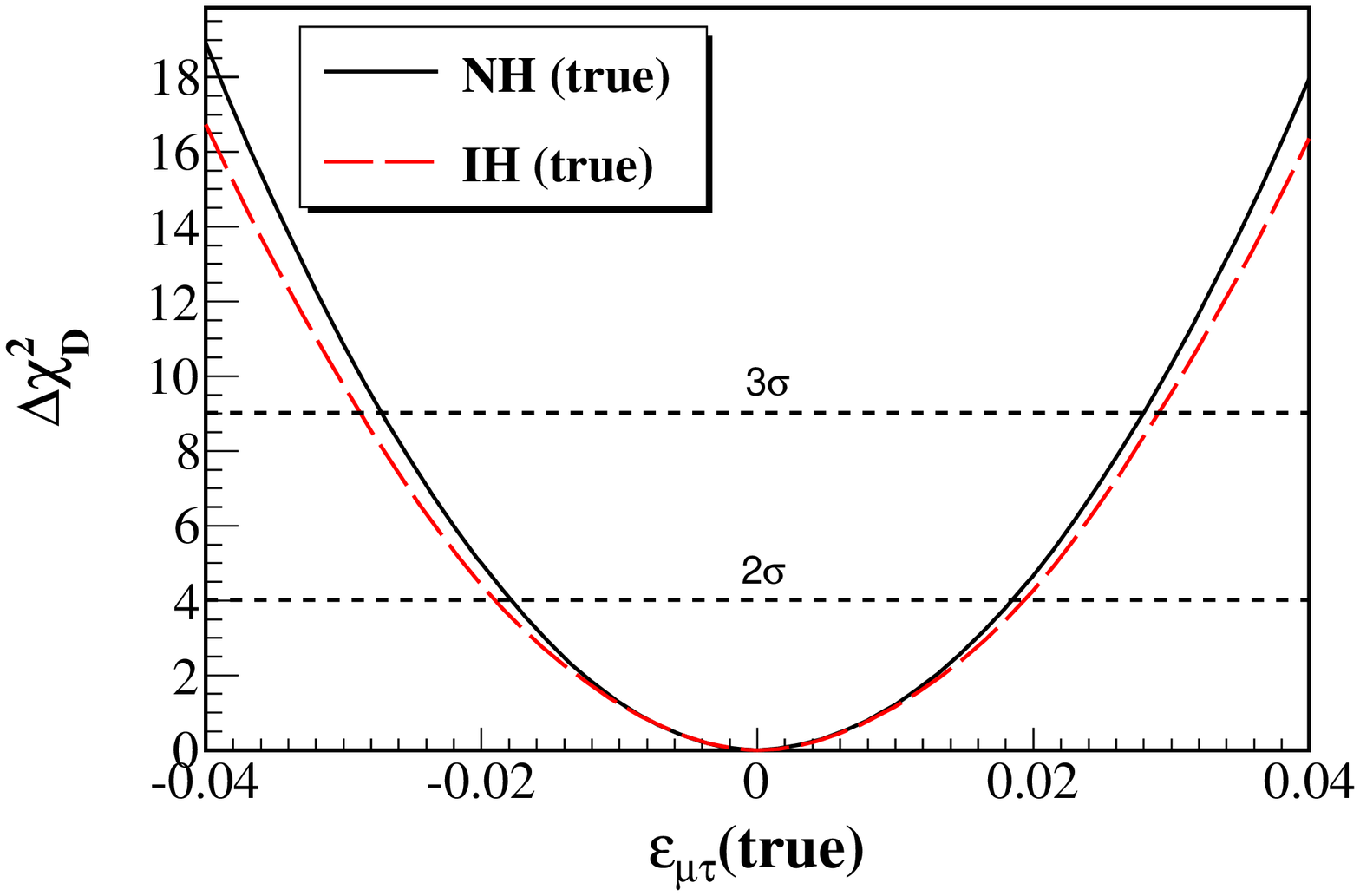}
\includegraphics[width=0.495\textwidth]{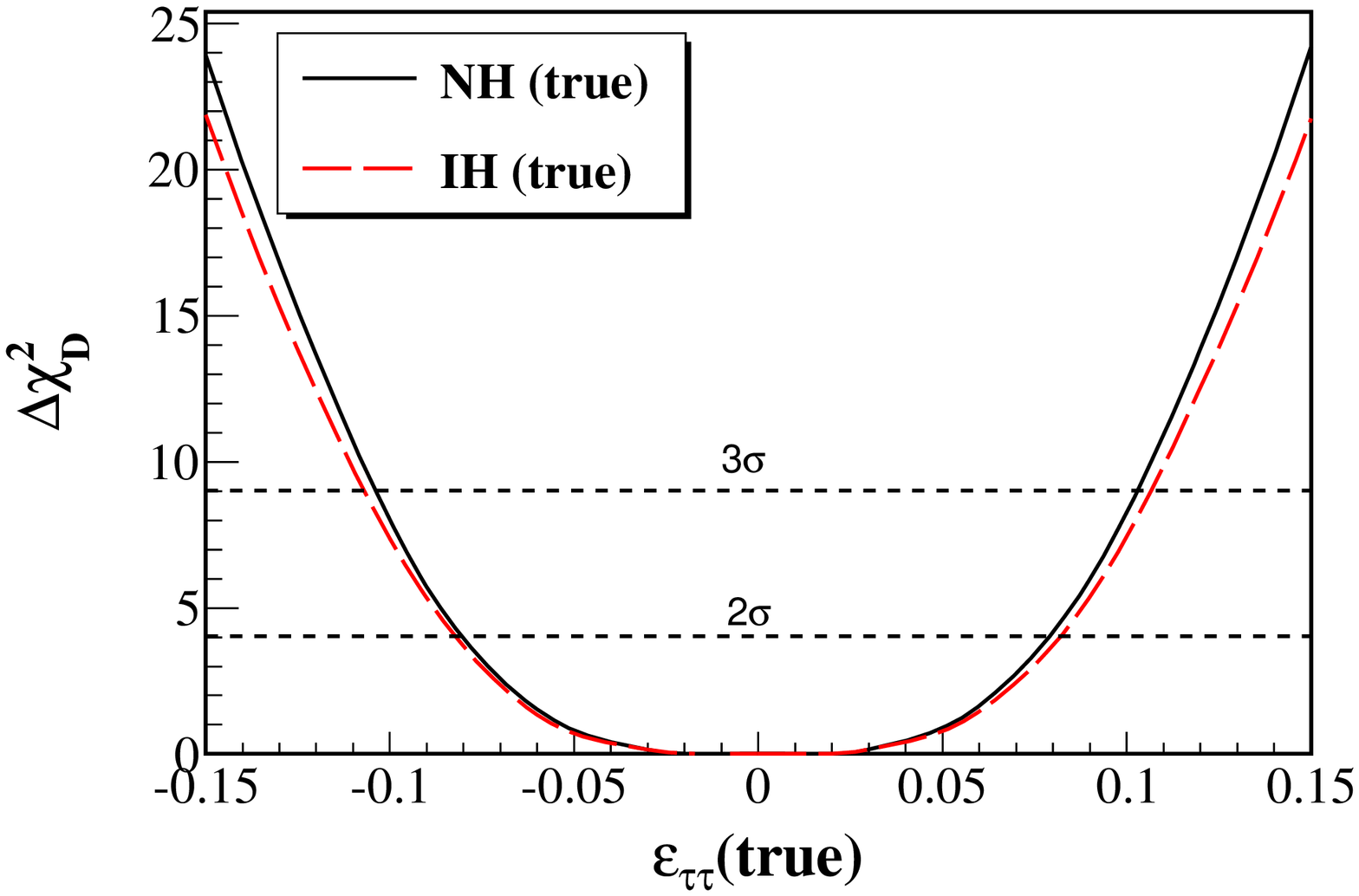}
\caption{
The $\Delta \chi^2_{\rm D}$, giving the discovery potential of 10 years of 
ICAL data in observing NSI, as a function of the true value of NSI parameters. We keep only one NSI 
parameter to be non-zero at a time, while others are set to zero. 
The $\Delta \chi^2_{\rm D}$ has been marginalised over the oscillation parameters as explained 
in the text.
}
\label{fig:discovery}
\end{center}
\end{figure}

In the previous section, we looked at how well ICAL will be able to constrain NSI 
parameters if its data were consistent with just standard oscillations. Next, we take 
the complementary view and ask ourselves that if NSI parameters were indeed 
non-zero, at what C.L.~would ICAL be able to tell them apart from standard oscillations. 
In other words, we are looking for the limiting true values of the NSI 
parameters above which the data at ICAL would be a signal for NSI 
at a certain C.L.
For that, we now consider data for various (assumed) true values of the NSI 
parameters and fit them with a predicted event spectrum corresponding to standard 
oscillations and compute the resultant $\Delta \chi^2_{\rm D}$. 
We present this in 
Fig.~\ref{fig:discovery}. For simplicity, we consider only one non-zero NSI 
parameter at a time in the data. We marginalise over the oscillation 
parameters $|\ma|$, $\sin^2\theta_{23}$, and $\sin^2\theta_{13}$ with priors 
imposed on each one of them as discussed before. The black solid curves 
correspond to the case for NH, while the red dashed curves are for IH. We keep 
the hierarchy to be the same in the theory as in the data. The figure shows the 
expected $\Delta \chi^2_{\rm D}$ for the discovery of each of the NSI parameters 
$\epsilon_{e\mu}$ (top-left panel),
$\epsilon_{e\tau}$ (top-right panel),
$\epsilon_{\mu\tau}$ (bottom-left panel), and 
$\epsilon_{\tau\tau}$ (bottom-right panel). 
While the nature of the curves are very similar to the ones we had in 
Fig.~\ref{fig:sens}, the values of the $\Delta \chi^2_{\rm D}$ are different. 
With 500 kton-years of data, the ICAL experiment will be able to give a signal of NSI at 
the 90 \% (3$\sigma$) C.L.~for NH if
\be
 \epsem <  -0.116~(-0.28),  &&  \epsem > 0.105~(0.2) \,, \nn \\
  \epset < -0.12~(-0.29), &&   \epset >  0.102~(0.23) \,, \nn \\
 \epsmt < -0.015~(-0.027),  && \epsmt > 0.015~(0.028) \,, \nn \\
  \epstt <  -0.07~(-0.104),  &&  \epstt > 0.07~(0.103) \,. \nn 
\ee
The corresponding limiting values for IH are similar, as can be seen from the figure.

\subsection{\label{precision}Precision on NSI Parameters }
\begin{figure}[]
\begin{center}
\includegraphics[width=0.32\textwidth]{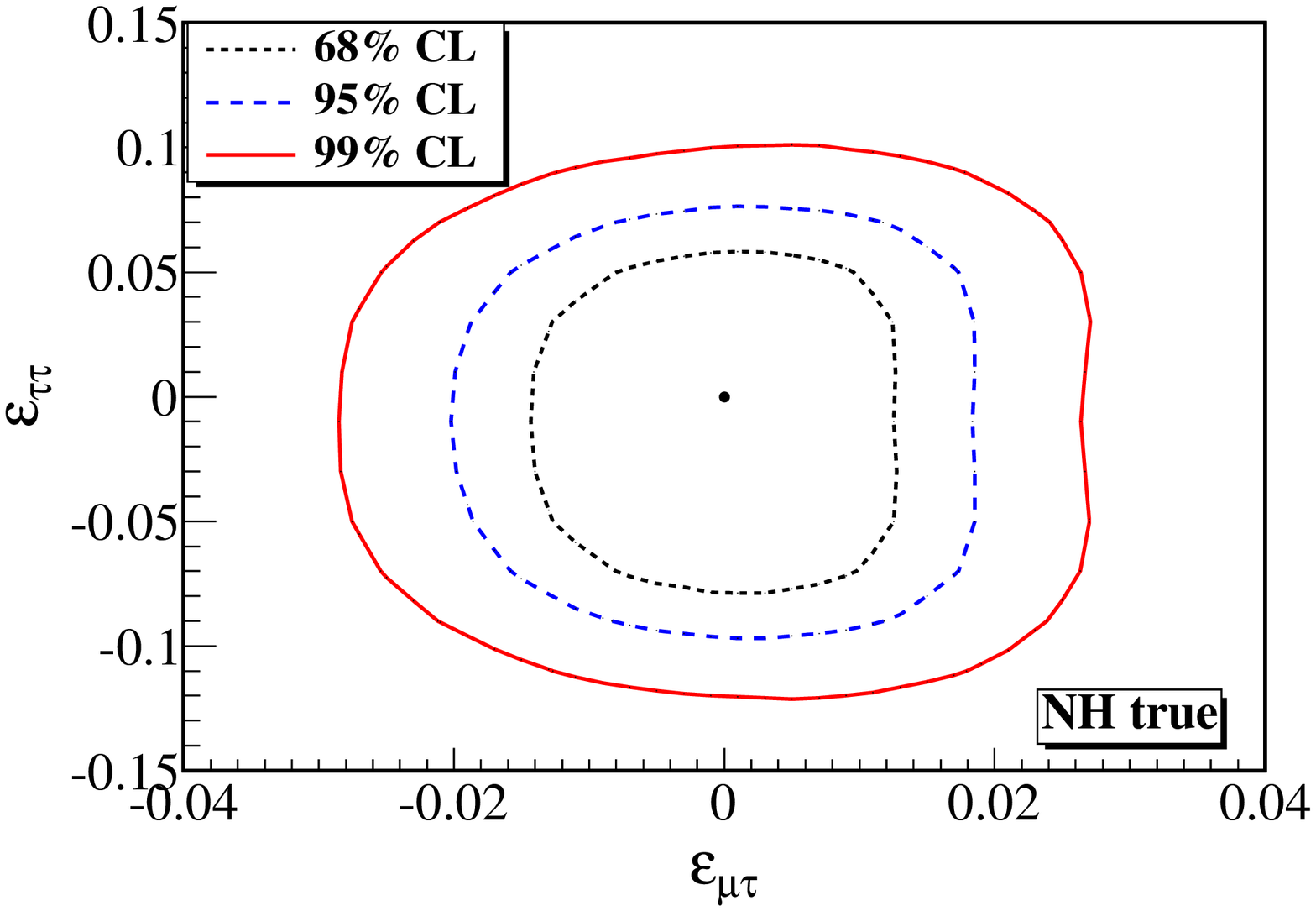}
\includegraphics[width=0.32\textwidth]{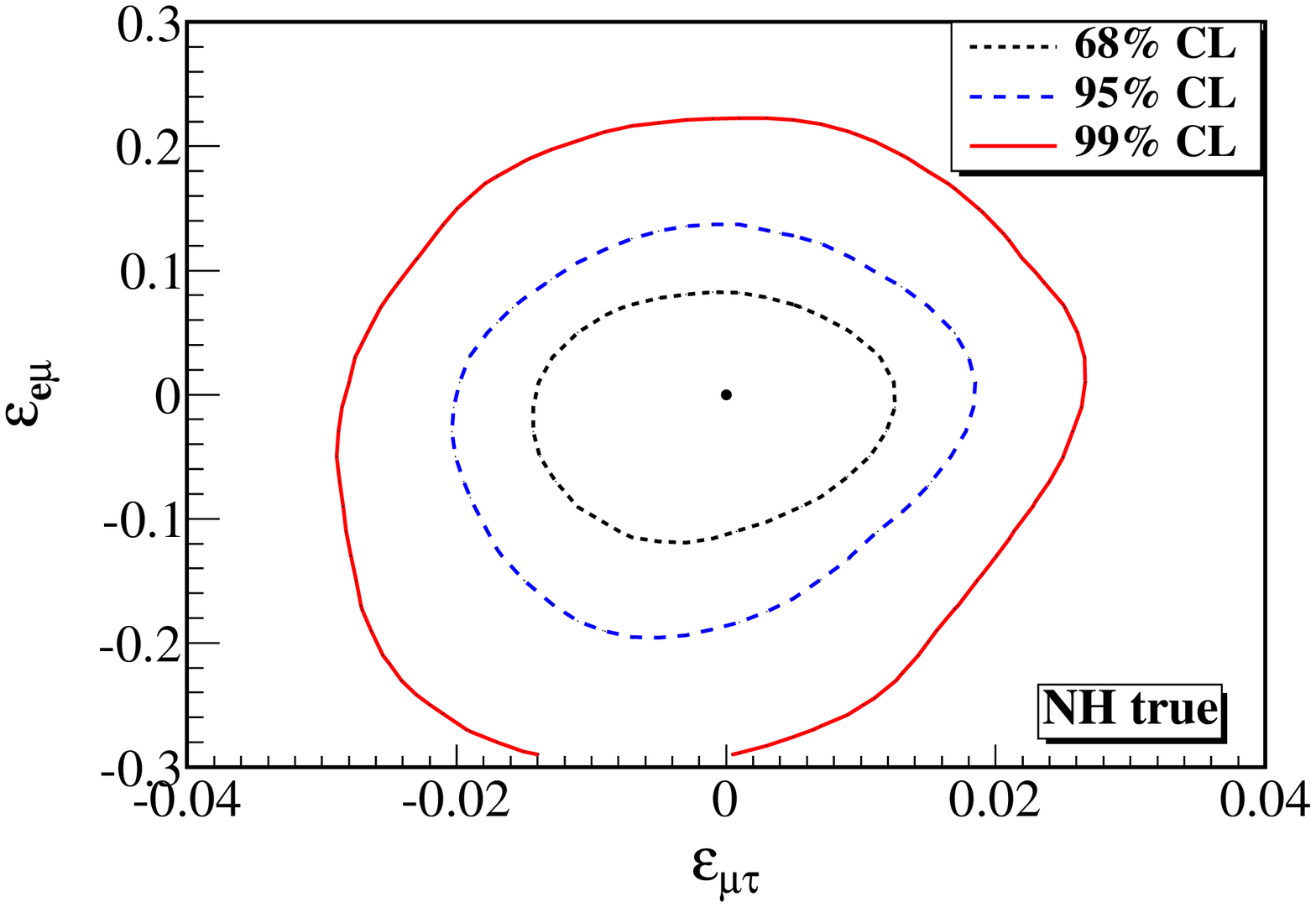}
\includegraphics[width=0.32\textwidth]{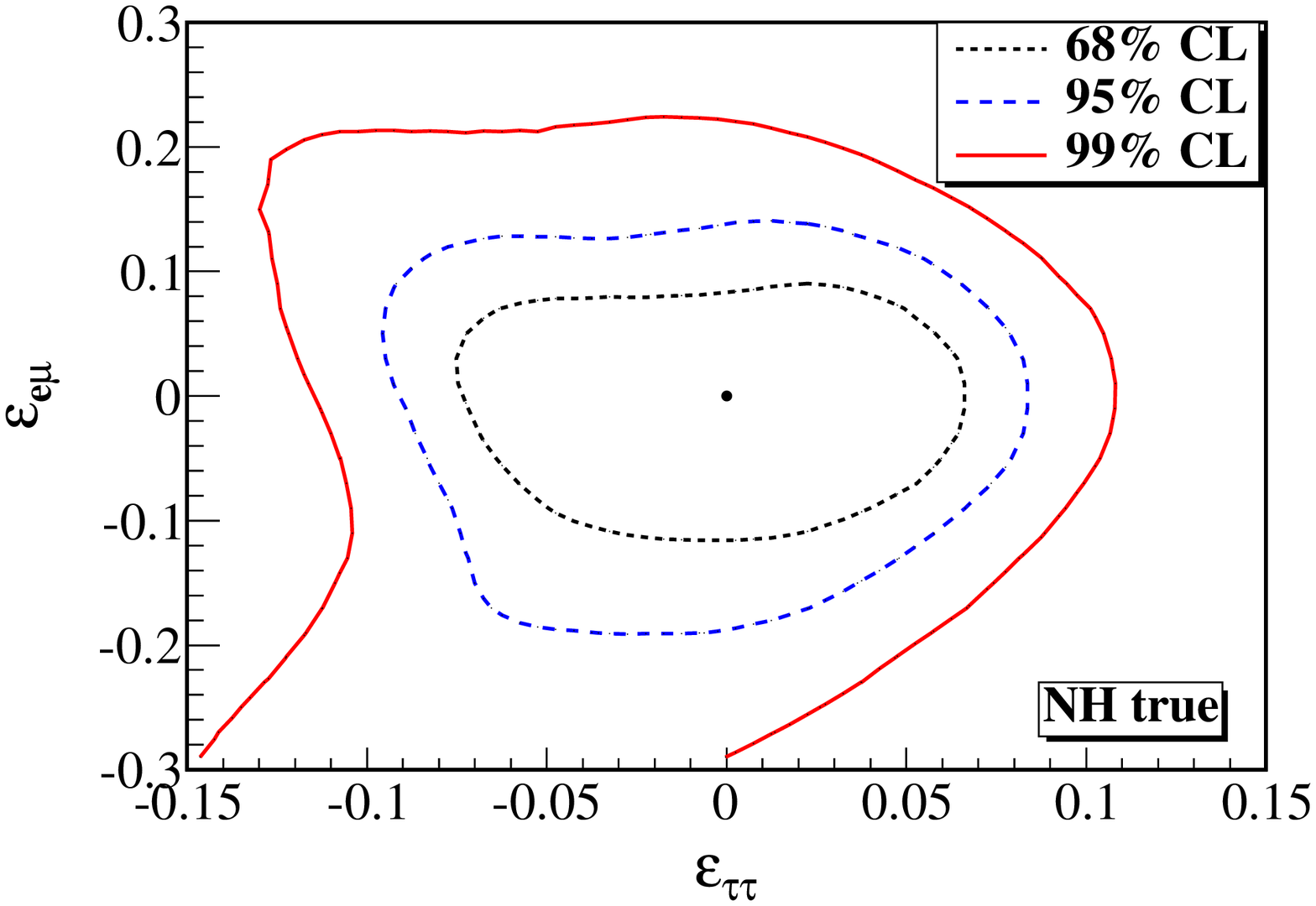}
\includegraphics[width=0.32\textwidth]{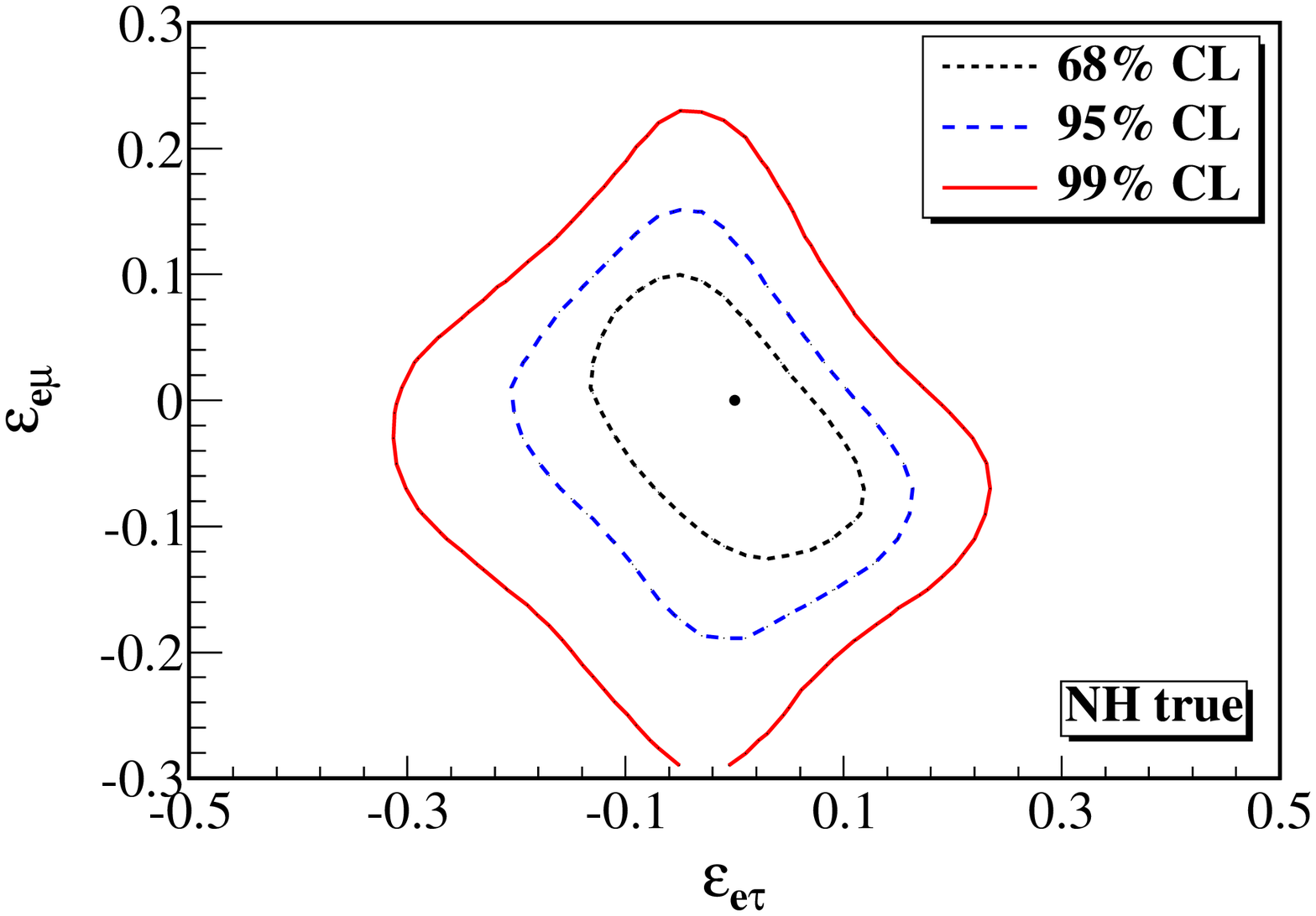}
\includegraphics[width=0.32\textwidth]{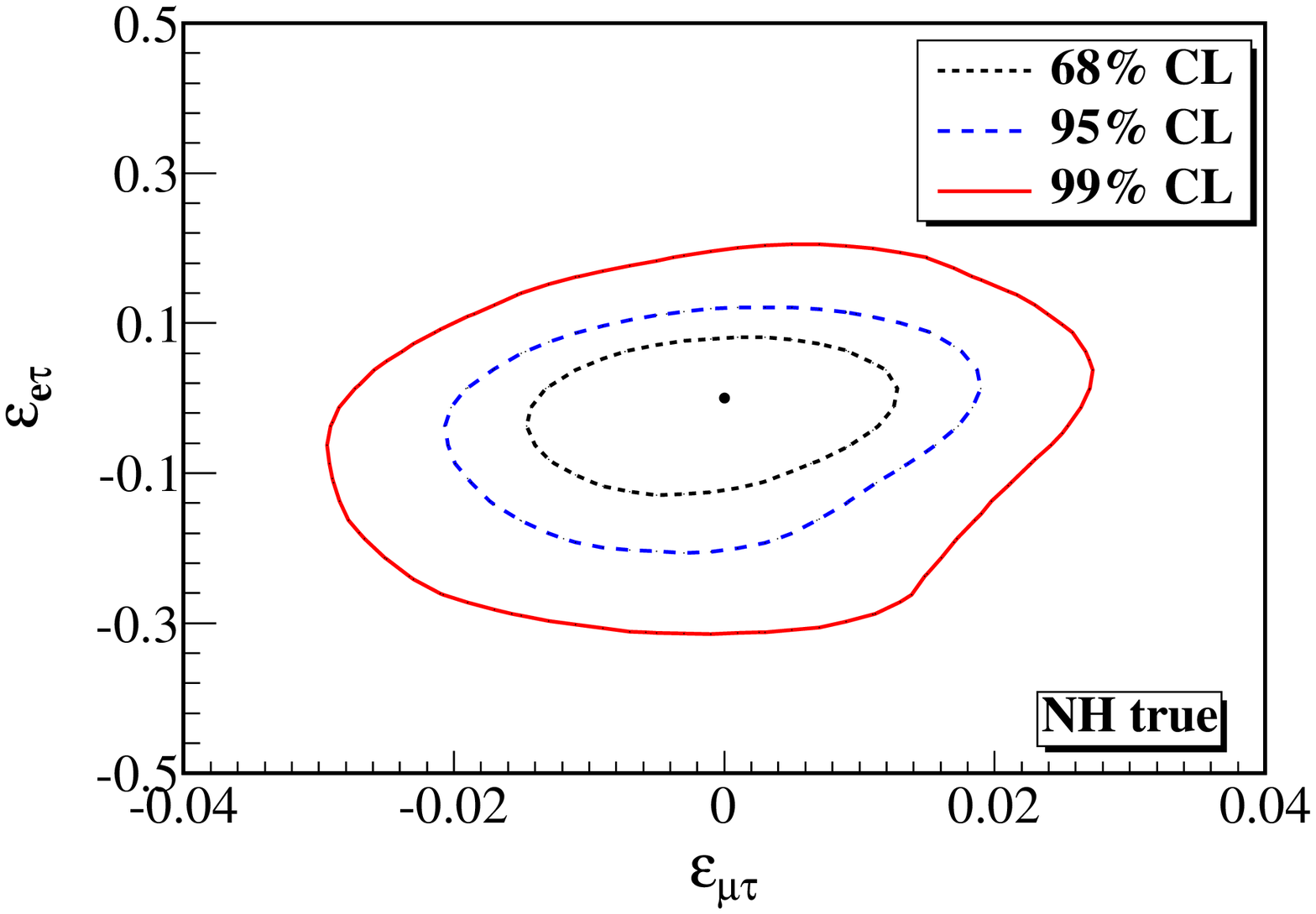}
\includegraphics[width=0.32\textwidth]{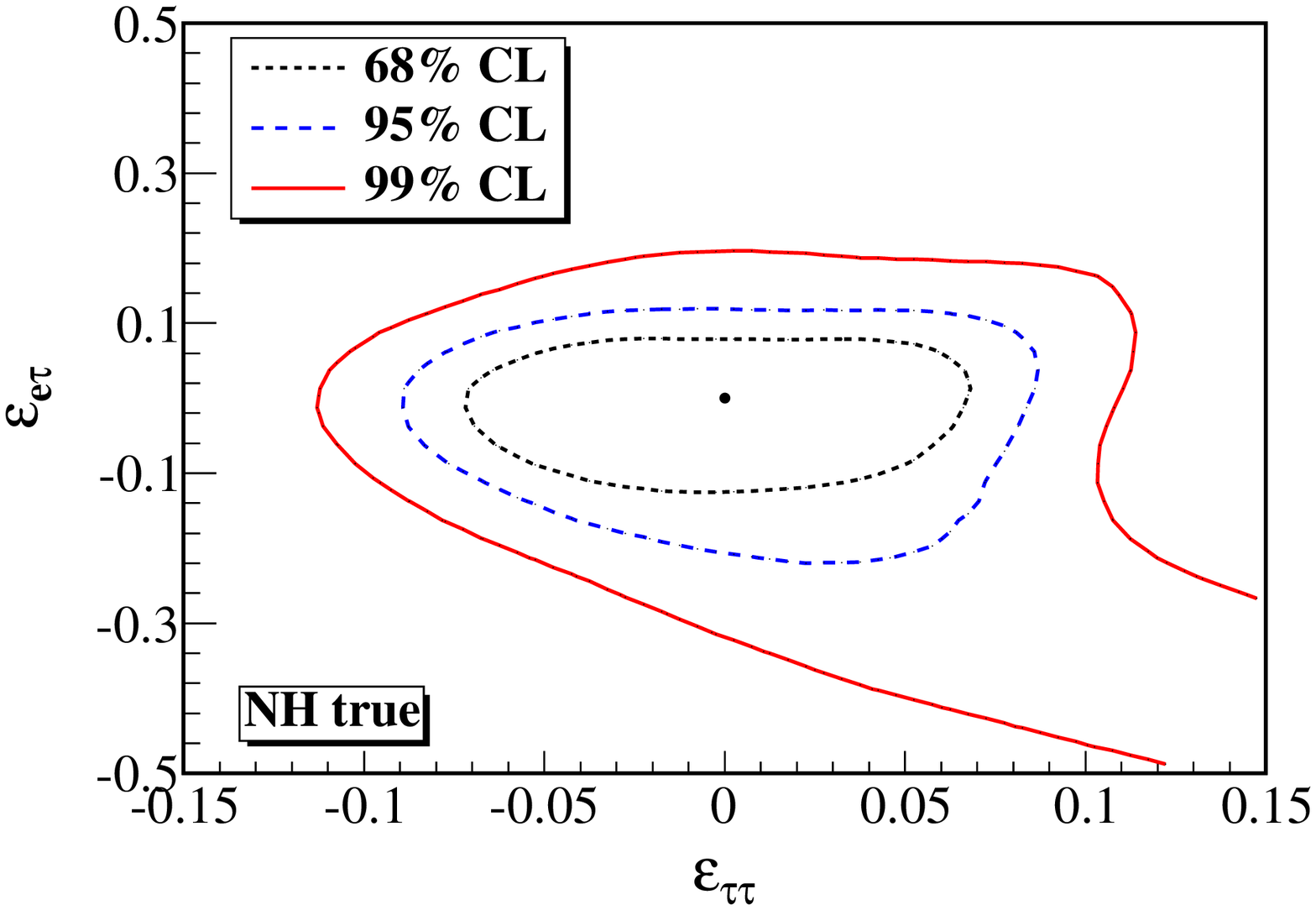}
\caption{The expected C.L.~contours in the given NSI parameter plane. The other 
NSI parameters are set to zero. The NH 
has been assumed to be true. The black dots show 
the points where the data were generated, which are for no NSI in this case. 
}
\label{fig:precis1}
\end{center}
\end{figure}


\begin{figure}[]
\begin{center}
\includegraphics[width=0.32\textwidth]{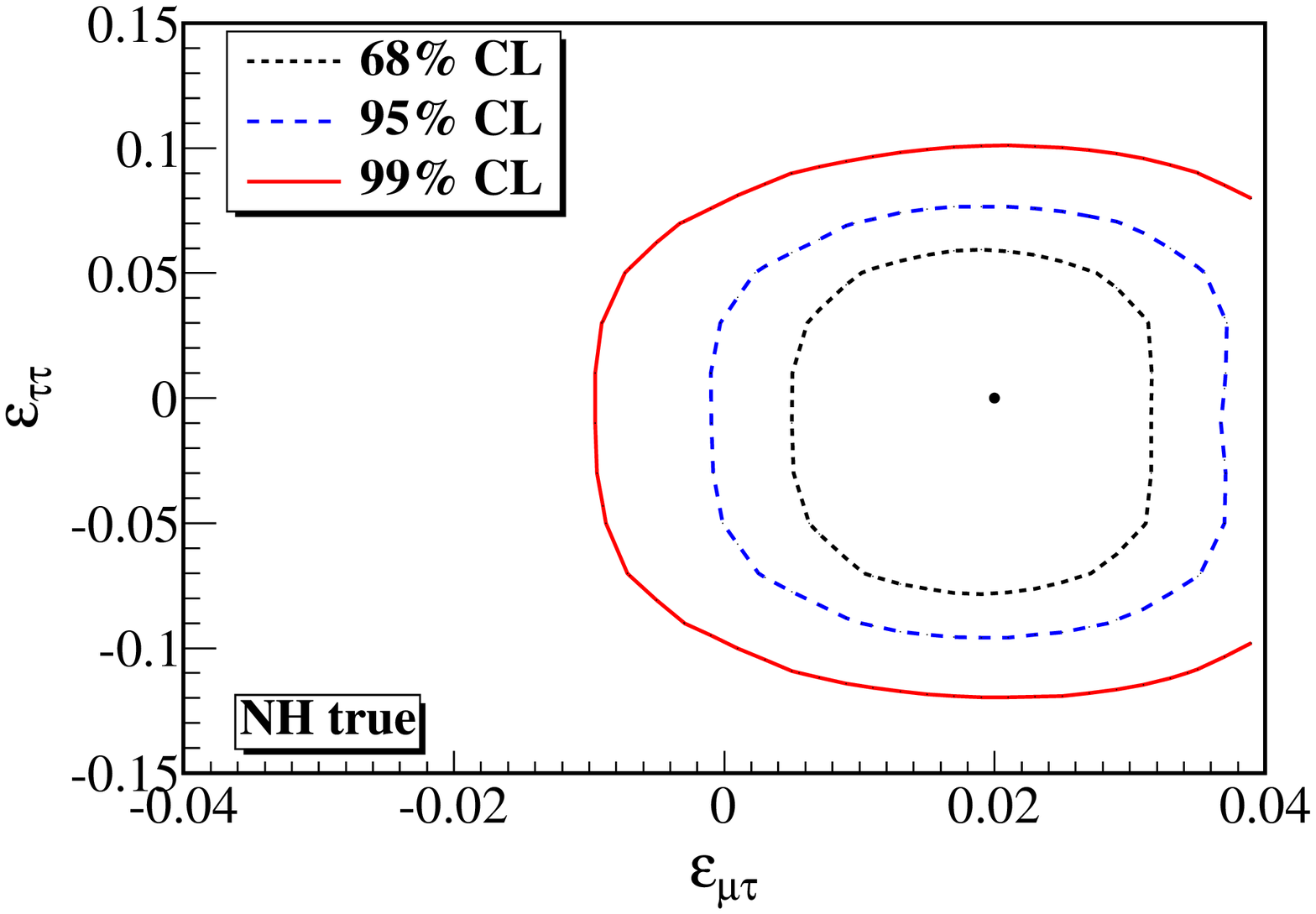}
\includegraphics[width=0.32\textwidth]{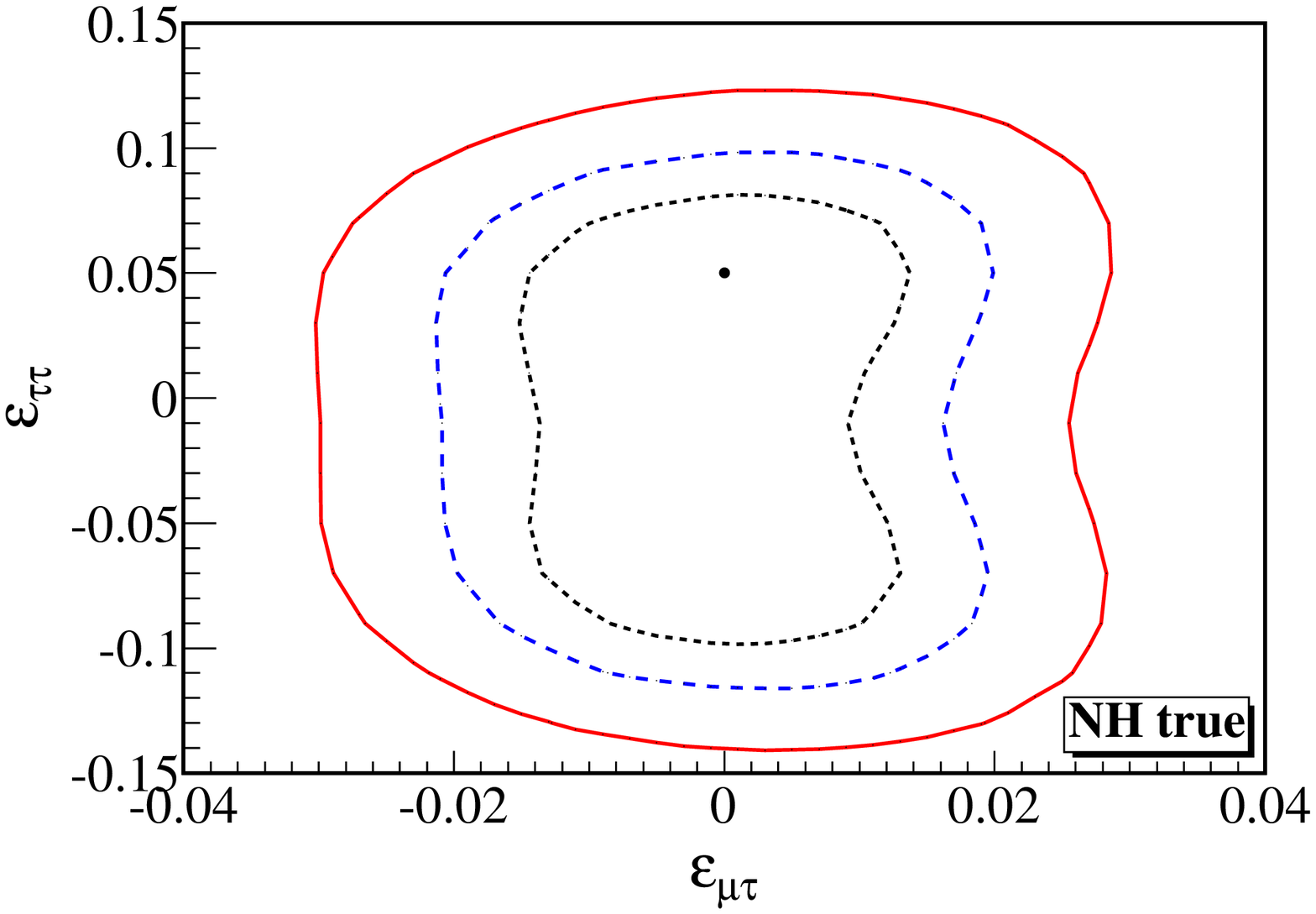}
\includegraphics[width=0.32\textwidth]{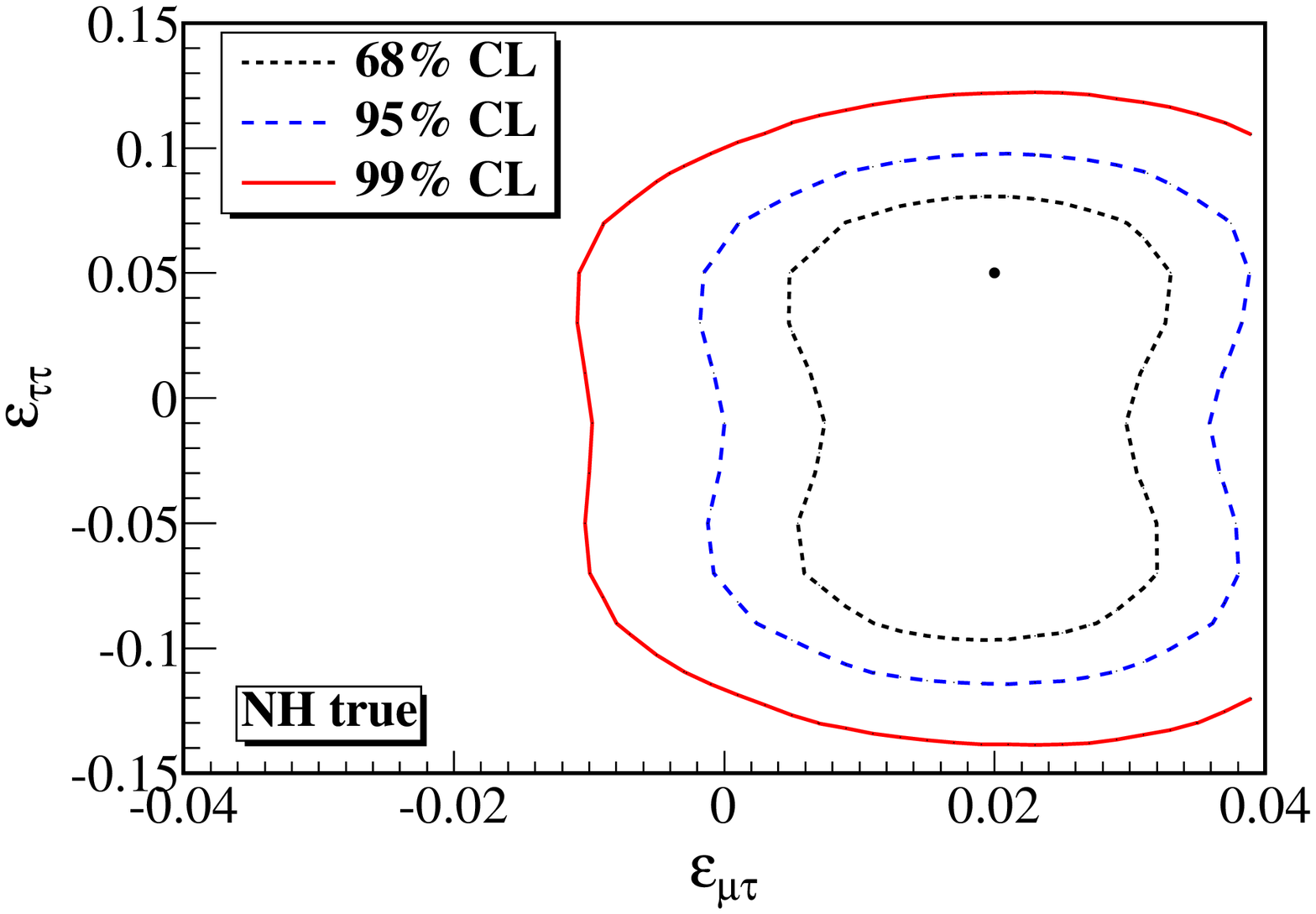}
\includegraphics[width=0.32\textwidth]{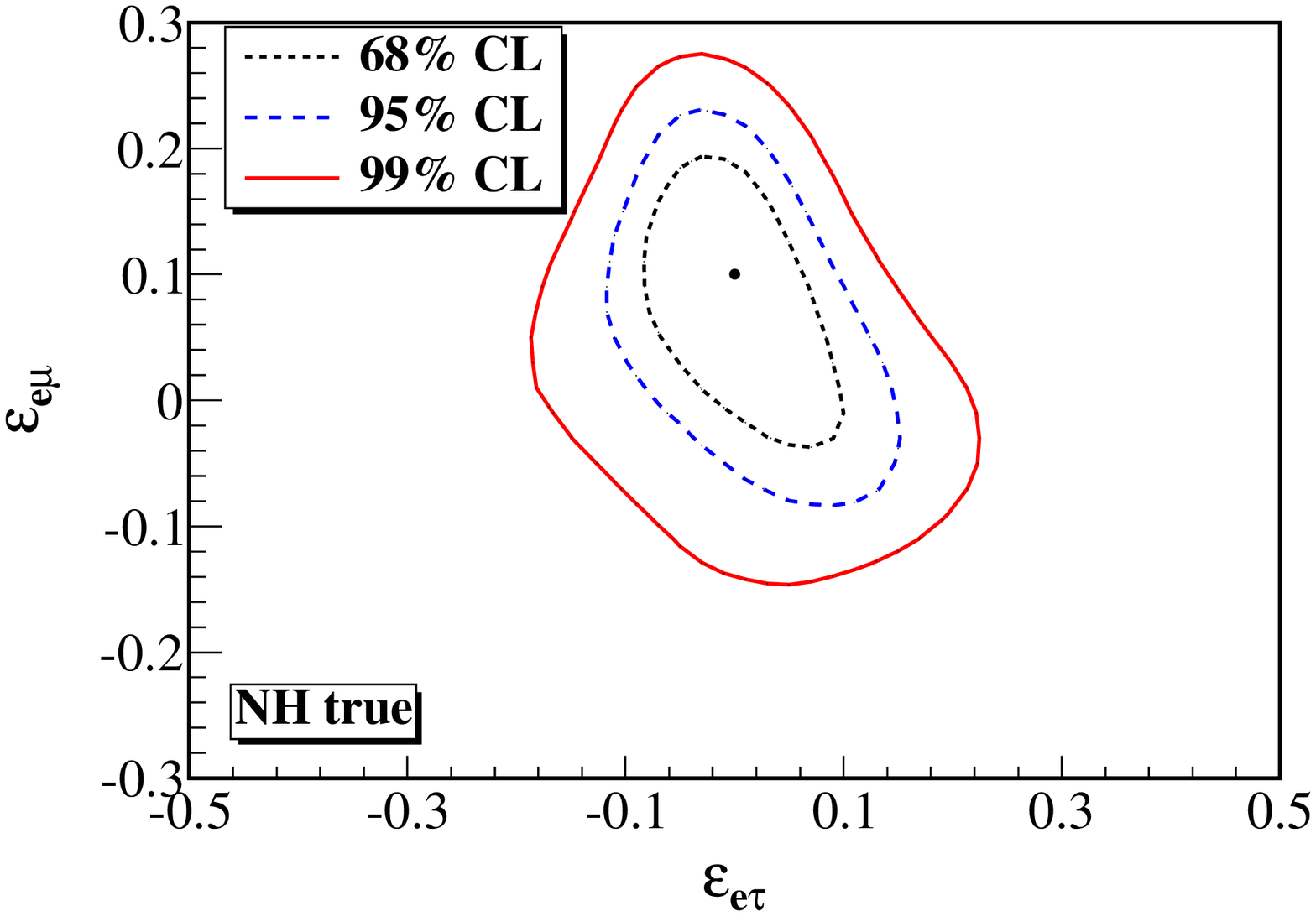}
\includegraphics[width=0.32\textwidth]{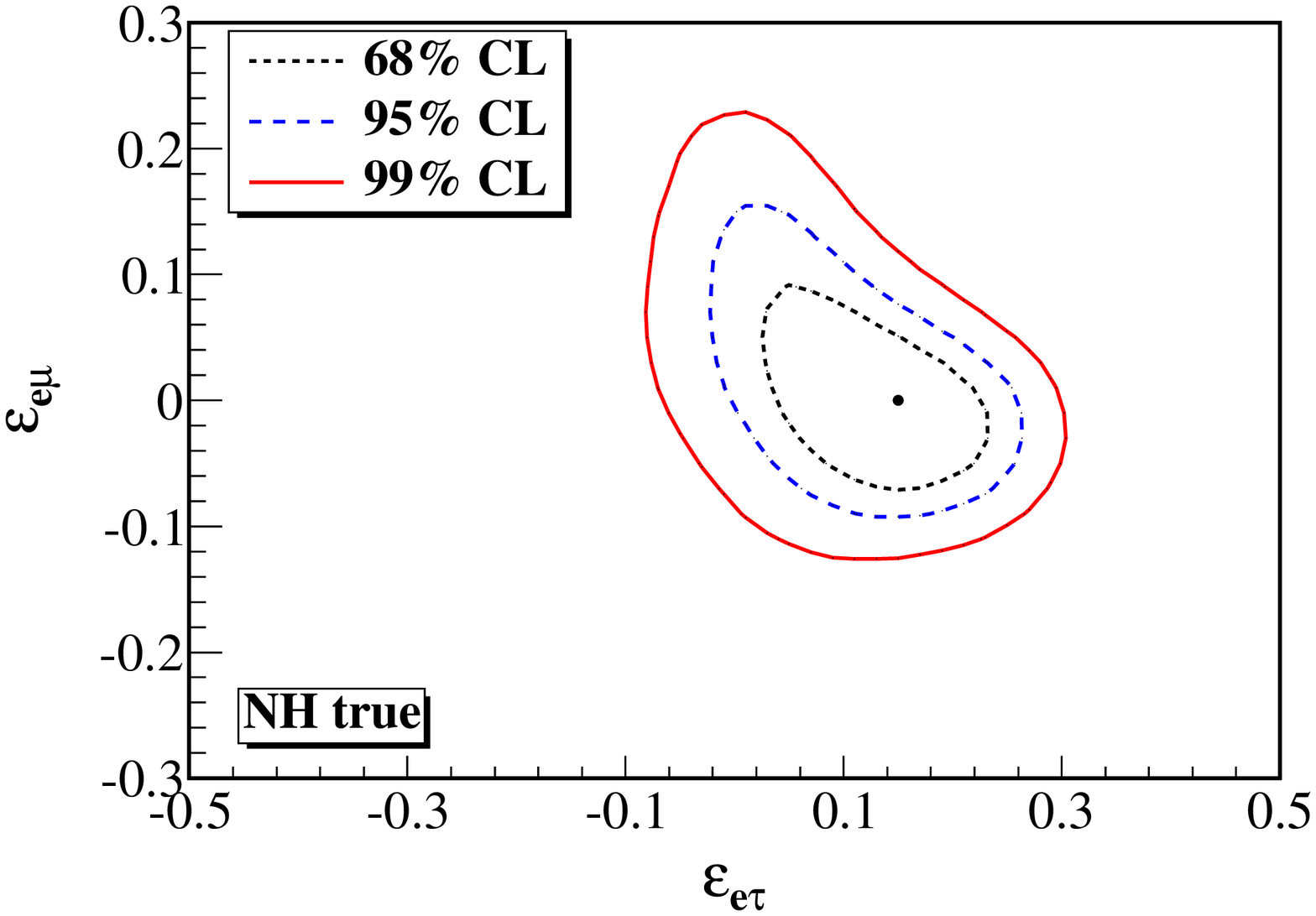}
\includegraphics[width=0.32\textwidth]{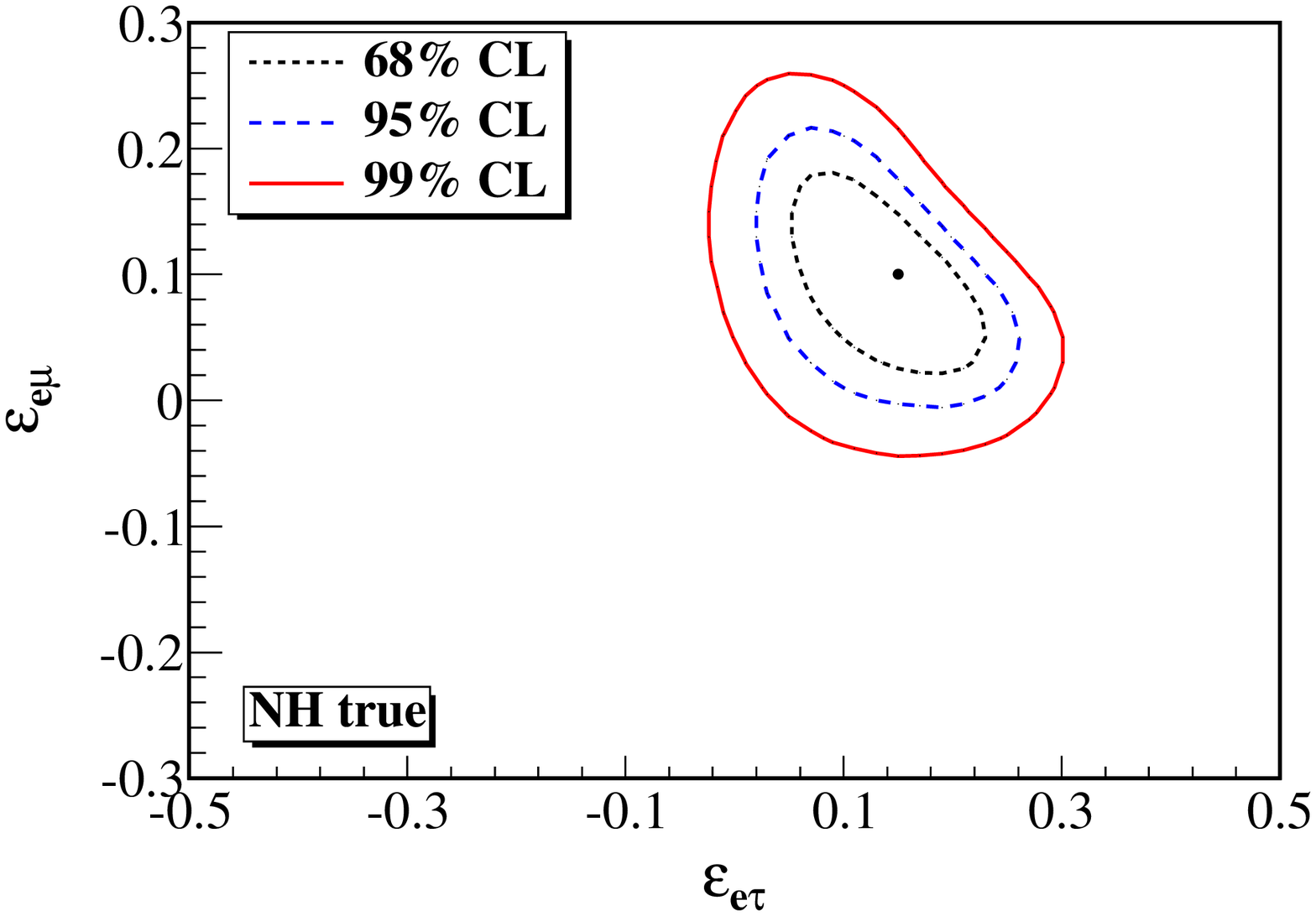}
\caption{The expected C.L.~contours in the given NSI parameter plane. 
The other 
NSI parameters are set to zero. 
The upper 
panels are drawn in the $\epsmt-\epstt$ plane, while the lower panels 
are drawn in the $\epset-\epsem$ plane. The black dots show 
the points where the data were generated.
}
\label{fig:precis3}
\end{center}
\end{figure}

In Figs.~\ref{fig:precis1}
and \ref{fig:precis3}, we show the 
projected C.L.~area in the NSI parameter space allowed after 
10 years of running of the ICAL experiment. In Fig.~\ref{fig:precis1}, 
we show these for the case where we assume that there are 
no NSI, or in other words, when the assumed true values of the 
NSI parameters are taken as zero, shown by the black dots in the 
figure. The black dotted, blue dashed, and red solid contours show the 
68~\%, 95~\% and 99~\% C.L.~in two-dimensional NSI planes. 
The contours are marginalised over the oscillation 
parameters after including priors that are described above. 
The NSI parameters other than the ones appearing in the two-dimensional 
plane are kept fixed at zero. NH is assumed for all plots. The 
corresponding contours for the IH case are very similar and we 
do not repeat them for brevity. 

In Fig.~\ref{fig:precis1},
we show the C.L.~contours for the case where we assume the 
true values of the NSI parameters to be non-zero. The upper 
panels show the C.L.~in the $\epsmt$-$\epstt$ plane when the 
true values of ($\epsmt,\epstt$)  are taken as (0.02,0), (0,0.075), and 
(0.02,0.075) for the left, middle, and right panels, respectively. 
These assumed true points are shown by black dots in the plots. 
The values $\epsem$ and $\epset$ are assumed to be zero in both 
the data as well as in the fit, and we do not show the contours  
in planes involving these parameters. The lower panels are 
similar to the upper panels except that now we 
show the C.L.~in the $\epset$-$\epsem$ plane when the 
true values of ($\epset,\epsem$)  are taken as (0,0.1), (0.15,0), and 
(0.15,0.1) for the left, middle, and right panels, respectively. 
For these panels, the values $\epsmt$ and $\epstt$ are assumed to be zero in both 
the data as well as in the fit. Again, the figures are for NH, however, the ones 
for IH are similar and we do not present them for brevity.

\section{\label{conclusions}Conclusions}

The study of the physics potential of the ICAL detector at the proposed India-based 
Neutrino Observatory is underway. As a part of this on-going effort, we probe in 
this work the impact of NSI parameters on the expected signal at ICAL and the 
physics conclusions that one can draw out of it. The neutral-current NSI if present, 
could alter the propagation of atmospheric neutrinos inside the Earth matter 
changing the signal at ICAL. This change due to NSI can be used  
to study the NSI parameters. On the other hand, one needs to estimate how much 
the potential of ICAL to standard physics gets modified in the presence of NSI. In 
this work, we have taken both these considerations into account and studied the 
physics potential of ICAL in presence of NSI. 

Measurement of the neutrino mass hierarchy is the primary goal of the ICAL 
atmospheric neutrino experiment. We showed how the difference in the neutrino 
oscillation probabilities between NH and IH change in 
presence of NSI. We defined the relative probability difference $A_{\alpha\beta}^{\rm MH}$ 
for the oscillation channel $\nu_\alpha \to \nu_\beta$ and showed the oscillograms 
for $A_{\mu\mu}^{\rm MH}$ and $A_{e\mu}^{\rm MH}$, the two oscillation channels 
relevant for the atmospheric neutrinos in ICAL. These oscillograms (and all 
other results shown in this paper) were 
obtained from an exact numerical calculation of the 
three-generation neutrino oscillation probabilities using the PREM 
profile for the Earth matter density \cite{Dziewonski:1981xy}. 
The oscillograms show that the relative 
difference $A_{\mu\mu}^{\rm MH}$ changes significantly with $\epsmt$ 
compared to its Standard Model value, while $A_{e\mu}^{\rm MH}$ is seen to 
vary sharply with the values of $\epsem$ and $\epset$. The impact of the NSI 
parameter $\epstt$ is seen to be less important. 

We next simulated $\mu^-$ and $\mu^+$ events in the ICAL detector in presence of NSI
and defined a $\chi^2$ function, including energy and zenith angle correlated 
as well as uncorrelated systematic uncertainties, to give C.L.~predictions for the estimated 
sensitivity of ICAL.\footnote{It has been shown that the inclusion of hadron energy information in the 
analysis of ICAL data improves the 
mass hierarchy sensitivity of ICAL. The impact of the hadron energy information 
on the sensitivity of ICAL to NSI is being studied in an independent work \cite{nsihadron}.}
The $\chi^2$ is marginalised over the NSI parameters and 
the oscillation parameters $|\ma|$, $\sin^2\theta_{23}$, and $\sin^22\theta_{13}$ 
after putting priors on them. 
Using this we presented the change in $\Delta \chi^2_{\rm MH}$ 
if NSI was assumed to be a certain true value in Nature. We showed that the 
$\Delta \chi^2_{\rm MH}$ increases rapidly for $\epsem$(true)$>0$ and $\epset$(true)$ >0$, 
while it decreases for $\epsem$(true)$<0$ and $\epset$(true)$ <0$ 
compared to what we expect for standard oscillations. This behavior can be understood 
from the oscillograms we showed. The impact of the NSI parameter $\epstt$ 
is small, however, the $\Delta \chi^2_{\rm MH}$ could vary significantly with $\epsmt$.
However, if we allow for marginalisation over the oscillation (especially $|\ma|$) 
and NSI parameters, the $\Delta \chi^2_{\rm MH}$ comes to be around the value 
predicted by the Standard Model. 

We next showed the potential of ICAL in discovering or constraining NSI. If the case 
that ICAL was consistent with no NSI in the data, we presented the expected 
upper limit on the NSI parameters. At the 90~\% ($3\sigma$) C.L.~we have
for the NH the limits 
\be
-0.119~(-0.3)& <  \epsem <  &0.102~(0.2) \,, \nn \\
-0.127~(-0.27) &<  \epset <  &0.1~(0.23) \,, \nn \\
-0.015~(-0.027) &<  \epsmt <  &0.015~(0.027) \,, \nn \\
-0.073~(-0.109)& <  \epstt <  &0.073~(0.109) \,. \nn 
\ee
The limits for IH are similar. 
Compared to the current 90\% C.L.~bounds given in 
Eqs. (\ref{eq:bound}) and (\ref{eq:skbound}) the expected bounds from ICAL are promising. 
We next considered the case where the data at ICAL is consistent with NSI and we 
gave the expected statistical significance with which ICAL will rule out 
the theory with no NSI. We calculated the range of the NSI parameters that would 
lead to 90\% ($3\sigma$) C.L.~discovery of NSI at ICAL. Finally, we presented the 
C.L.~contours in the two-parameter NSI planes, for different choices of 
true values of the NSI parameters.

\begin{acknowledgments}
This work is a part of the ongoing effort of INO-ICAL collaboration to study various physics potential of the proposed INO-ICAL detector. Many members of the collaboration have contributed for the completion of this work. In particular, we thank S. Agarwalla, A. Chatterjee, A. Dighe, S. Goswami, A. Khatun, and T. Thakore for discussions. We acknowledge N. Sinha for discussions during the early 
stages of this work. 
S.C.~acknowledges support from the Neutrino Project under the XII plan of Harish-Chandra Research Institute and partial support from the European Union FP7 ITN INVISIBLES (Marie Curie Actions, PITN-GA-2011-289442).
\end{acknowledgments}

\bibliography{references}

\end{document}